\documentclass[]{aa} 
\usepackage{graphicx}
\usepackage{txfonts}
\usepackage{natbib}
\bibpunct{(}{)}{;}{a}{}{,}
\begin{document}
\title{Multi-wavelength observations of the young binary system Haro~6-10: The case of misaligned discs      
\thanks{Based on observations collected at
   the European Southern Observatory, Paranal, Chile (Proposal IDs:
   070.C-0544(A); 072.C-0321(A); 274.C-5032(A); 076.C-0200(A))}
}


   \author{V. Roccatagliata\inst{1,2}
	   \and
            Th. Ratzka\inst{3}
	   \and
 	   Th. Henning\inst{2}
	   \and
            S. Wolf\inst{4}
	   \and
           Ch. Leinert\inst{2}
	   \and
 	   J. Bouwman\inst{2}
}
  \offprints{V. Roccatagliata: rocca@stsci.edu}

   \institute{Space Telescope Science Institute, 3700 San Martin Drive, Baltimore, MD 21218, USA
              \email{rocca@stsci.edu}
	      \and 
              Max-Planck-Institut f\"ur Astronomie (MPIA), K\"onigstuhl 17, D-69117 Heidelberg, Germany
              \and
              Universit{\"a}ts-Sternwarte M{\"u}nchen, Ludwig-Maximilians-Universit{\"a}t, Scheinerstr. 1, 81679, M{\"u}nchen, Germany
	      \and
              University of Kiel, Institute of Theoretical Physics and Astrophysics, Leibnizstrasse 15, D-24098 Kiel, Germany
            }

   \date{Received March 1, 2011; Accepted July 15, 2011}

 
  \abstract
   {We present a multi-wavelength, high-resolution observational survey of the young binary system Haro 6-10 (GV~Tau, IRAS 04263+2426), which is harbouring one of the few known infrared companions. 
   }
   {The primary goal of this project is to determine the physical and geometrical properties of the circumstellar and circumbinary material in the Haro~6-10 system.
     }
   {High-resolution optical (HST/WFPC2) and near-infrared (VLT/NACO) images in different bands were analysed to investigate the large-scale structures of the material around the binary. Mid-infrared interferometry (VLTI/MIDI) and spectroscopy (TIMMI2 at the 3.6m ESO telescope) were carried out to determine the structure and optical depth of the circumstellar material around the individual components.
     }
   {The multi-wavelength observations suggest that both components of the binary system Haro~6-10 are embedded in a common envelope. The measured extinction indicates a dust composition of the envelope similar to that of the interstellar medium. Each component of the system has a circumstellar disc-like structure typical of young stars. The discs are highly misaligned: the northern component is seen almost edge-on and the southern component is an almost face-on disc. 
}
   {
	The two main formation scenarios of binary systems with misaligned discs are the gravitational capture of a passing object in a dense environment, and the fragmentation of the collapsing molecular cloud. Given the low-density environment of the Taurus-Aurigae star-forming region, the first scenario is unlikely for Haro~6-10. 
	The binary system most probably formed 
	via fragmentation of two different parts of the collapsing molecular cloud combined with other dynamical processes related to the cloud and/or the protostars. This can be the explanation also for other binary systems with an infrared companion.
}

   \titlerunning{Multi-wavelengths observations of Haro~6-10} 
   \authorrunning{Roccatagliata et al.}
   \maketitle
%

\section{Introduction}

The formation of binary and multiple systems seems to be a common phenomenon in star formation \citep[e.g.,][]{Mathieuetal2000}. Surveys of young star-forming regions with high spatial resolution allow a statistical description of the binary parameters and a comparison between regions of various ages and with different environments \citep[e.g.,][]{Leinertetal1993, Ducheneetal2007}. Binarity has been studied also in young stellar clusters of various ages to investigate its relation to the evolution of protoplanetary discs \citep[e.g.,][]{ClarkePringle1991a, Bouwmanetal2006}.\\
Different scenarios have been proposed to explain the formation of binary systems. The standard scenario supports the hypothesis that a binary or multiple system forms when the core of a molecular cloud fragments during its gravitational collapse. The fragmentation process can be divided into two main classes: direct  \citep[e.g.,][]{BossBodenheimer1979, BateBurkert1997} and rotational fragmentation \citep[e.g.,][]{Bonnell1994, BonnellBate1994a, BonnellBate1994b, Burkertetal1997}. While the direct fragmentation strongly depends on the initial density distribution in the molecular cloud, this is not the case for rotational fragmentation, which is caused by asymmetric instabilities in a rotating disc or ring. If the proto-binary accretes (only) gas with low angular momentum, a disc is formed only around the primary star while the secondary may have none. If instead the secondary developed its own disc, also the primary will have a disc \citep{BateBonnell1997}.

According to the two different scenarios of binary formation, the resulting circumstellar discs may have different inclinations. When the binary is formed via rotational fragmentation the discs are expected to be preferentially aligned. If, however, the stellar components and their own discs form directly from a collapsing core \citep[`prompt initial   fragmentation';][]{Pringle1989}, this can lead to two misaligned discs \citep{Bateetal2000}. Under more complicated initial conditions even a small cluster can be formed from such a core. Moreover, interactions in these clusters between stars with discs can cause a dissipative encounter \citep[e.g.,][]{ClarkePringle1991a}. This may lead to the formation of binary or multiple systems via capture and thus to binaries with misaligned discs. In dense star-forming regions, a massive accretion disc may even cause the capture of a passing star \citep[][]{ClarkePringle1991b}. On the other hand, strong tidal interactions can lead to aligned discs.  Another possibility to have misaligned discs is that the gravitational interaction of a passing object causes the disc's tilt (and hence also the direction of a possible jet) in a binary system \citep[][]{Bateetal2000}. \cite{Heller1993} estimated that in one million years, in a small cluster, a few percent of the stellar systems can experience a tilt of a few degrees during the star encounters. 
The effect of stellar encounters in different cluster environments has been investigated by \citet[][]{Olczaketal2010}. They found that the encounter rate  remains unaffected by the size of the stellar population, while it strongly depends on the cluster density: in clusters less dense and less massive than the ONC, massive stars dominate the encounter-induced disc-mass loss, whereas in denser and more massive clusters the disc-mass removal is controlled by low-mass stars. 

In systems near the end of their accretion phase, the infall of a small amount of material with a different angular momentum to the orbit can cause a misalignment of discs originally aligned \citep[e.g.,][]{Bateetal2000}. 

Therefore, from the alignment of circumstellar discs it is possible to understand the dynamics that most likely took place during the star-formation process.

Studying disc alignment requires observations of protoplanetary discs around young binary stars with high spatial resolution. Previous studies determined the inclination by means of polarimetry \citep[e.g.,][]{Moninetal1998, Wolfetal2001, Moninetal2006}, orientation of the jets \citep[e.g.,][]{Eisloffeletal1996, Davisetal1997}, and direct observations of the discs \citep[e.g.,][]{Koresko1998, Stapelfeldtetal1998a}. Another technique, that allows to directly measure size, inclination and orientation of the inner parts of protoplanetary discs is long-baseline interferometry at infrared wavelengths. A recent interferometric study of T Tau showed that the three discs in this triple system are misaligned \citep{Ratzkaetal2009}. Therefore, the formation process of T Tau probably has been highly dynamic.

We present in this paper a multi-wavelength, high-resolution observational survey of the young binary system Haro 6-10 (GV~Tau, IRAS 04263+2426). Haro 6-10 is a T~Tauri binary system residing within the L-1524 molecular cloud. It is composed of an optically visible source, Haro 6-10~S, and an infrared companion (IRC), Haro~6-10~N. With an angular separation of 1.\arcsec2, Haro 6-10 N is actually located $\sim$160~AU north of Haro 6-10~S at the distance of the Taurus-Auruga star-forming region. The binary system was firstly resolved by \citet{LeinertHaas1989} using speckle interferometry. Haro~6-10~N has a very red spectral energy distribution and is brighter than Haro~6-10~S at wavelengths longer than 4~$\mu$m. The binary system is variable, particularly in the near-IR, on a timescale as short as a month \citep{Leinertetal2001}. \citet{Leinertetal2001} suggest that the variability of Haro~6-10~S is associated to inhomogeneities in the circumstellar material, while the variable behaviour of the northern component is likely associated to a variable accretion mechanism. The first resolved images in the near-infrared of this system, obtained by \citet{Menardetal1993}, suggest the presence of a flat circumbinary envelope, or disc.

The infrared spectrum of Haro 6-10 between 8 and 13~$\mu$m has been interpreted by \cite{Hanneretal1998} by an optically thick plus an optically thin emission, where the optically thin emission could be associated to an envelope, while the optically thick emission is associated to a disc, to the star itself, or to a featureless dust continuum. The Spitzer/IRS spectrum \citep[between 5-36~$\mu$m;][]{Furlanetal2008} shows a broad silicate absorption feature at 10~$\mu$m, but the spatial resolution of the instrument is too low to resolve the binary.

\citet{Reipurthetal2004} obtained high-resolution 3.6 cm observations with the VLA. In the 3.6\,cm map the two components of the binary are spatially resolved. At this wavelength the radio continuum emission originates from shock-ionised gas very close to the outflow exciting source \citep[e.g.,][]{Angladaetal1998}. \citet{Reipurthetal2004} suggest that Haro~6-10~S is the driving source of the large outflow activity seen in the near-infrared. 
HCN and C$_2$H$_2$ in absorption towards Haro~6-10~N were firstly detected by 
  \citet{Gibbetal2007} in the near-infrared using NIRSPEC at the 
  {\em Keck Telescope} and one year later by \citet{Doppmannetal2008}, using the same instrument.

In this paper, we are using a multi-wavelength approach, including different techniques, to characterise the binary system Haro 6-10. This paper is organised as follows: in Section~\ref{obs} we present the archival optical and near-infrared photometric observations, and our mid-infrared spectroscopic and interferometric observations. The data reduction and the results are described in Section~\ref{datared}. The analysis and the discussion of the results are presented in Sections~\ref{analysis} and \ref{discussion}, respectively. Conclusions are drawn in Section~\ref{conclusions}.

\begin{table}
\caption{Log of TIMMI~2 observations of Haro~6-10 and of the
  spectro-photometric standard stars.}
\label{timmi2}      
\centering               
\begin{tabular}{c c c c}  
\hline\hline                
\noalign{\smallskip}
Date       & UT    & Objects  & Airmass\\
\noalign{\smallskip}
\hline                
\noalign{\smallskip}
2002 Feb.~06 & 00:10 & HD32887  & 1.02\\
 	 & 00:41 & Haro~6-10    & 1.71\\
	 & 00:53 & Haro~6-10    & 1.73\\
	 & 01:41 & HD29139  & 1.56\\
	 & 07:46 & HD123139 & 1.11\\
\noalign{\smallskip}

2002 Dec.~26& 00:42 & HD29139  & 1.82\\
	& 01:35 & Haro~6-10    & 1.84\\
 	& 01:51 & Haro~6-10    & 1.78\\
& 05:00 & HD32887  & 1.07\\
\noalign{\smallskip}
\hline                        
\noalign{\smallskip}
\end{tabular}
\end{table}

\begin{table*}
\caption{Log of MIDI observations of Haro 6-10~N and S and calibrators observed. $^\star$: calibrators used to obtain the final calibrated visibilities and fluxes;  $^{\star\star}$: quality not good enough to be analysed}.          
\label{obs_gvtau}      
\centering              
\begin{tabular}{c c  r  c  c  c  c}  
\hline\hline                
\noalign{\smallskip}
&&&Airmass&Proj.~Baseline&Pos.~Angle&Notes\\
Date       &UT & Objects &&[m]&[deg] &\\
\noalign{\smallskip}
\hline                       
\noalign{\smallskip}
             &  &        &  &[UT2-UT3]&&\\
\noalign{\smallskip}
2004 Dec. 28 &00:18&  HD37160      &2.08& 27.61 & 14.61  & Calibrator$^\star$\\
	     &01:01&  Haro~6-10~N  & 1.73 &  28.43 & 48.67 & Science\\
	     &02:38&  Haro~6-10~S  &1.60& 38.13 & 52.15 & Science\\
             &03:10&  HD37160      &1.22& 39.98 & 43.62 & Calibrator$^\star$\\
             &04:08&  HD37160      &1.21& 43.61 & 46.24 & Calibrator$^\star$\\
             &05:26&  HD50778      &1.03& 46.43 & 42.72 & Calibrator\\
             &06:10&  HD94510      &1.43& 45.17 & 15.53 & Calibrator\\
             &07:02&  HD49161      &1.40& 46.61 & 45.68 & Calibrator\\
             &07:45&  HD94510      &1.25& 44.00 &32.44 & Calibrator\\
\noalign{\smallskip}
\noalign{\smallskip}
             &   &       &  &[UT3-UT4]&&\\
\noalign{\smallskip}
2005 Nov. 23 &04:17&  Haro~6-10~N$^{\star\star}$  &1.56& 62.13 & 106 & Science\\
             &05:22&  Haro~6-10~S  &1.54&  58.60 & 95.6  & Science\\
             &05:41&  Haro~6-10~N  &1.56&  56.80 & 92.6  & Science\\
             &06:09&  HD25604      &1.61& 50.53 & 85.4 & Calibrator$^\star$\\
             &07:49&  HD33042      &1.23& 61.60 & 116.5 & Calibrator\\
\noalign{\smallskip}

\hline                        
\end{tabular}
\end{table*}

\section{Observations}
\label{obs}
In this section we present the observations collected from the archives and our new TIMMI2 and VLT/MIDI observations of Haro~6-10~N and S. 

\subsection{Archival Optical and Near-infrared images}
\label{HST+NIRobs}
HST imaging observations of Haro~6-10 were obtained as part of the WFPC2 snapshot survey of nearby T~Tauri stars (HST 7387 program, PI: K. Stapelfeldt) in November~1998, using the HST PC1 (Planetary Camera 1) and the HST WF3 (Wide-Field Camera 3), which have a spatial scale of 0.05\arcsec px$^{-1}$ and 0.1\arcsec px$^{-1}$, respectively. The observations have been carried out with the F814W and F606W filters and consist of one exposure of 120 seconds with the F814W filter, and 2 exposures  with the F606W filter of 200 seconds and 600 seconds, respectively.
The calibrated data are available from the ESO/ST-ECF Science Archive Facility\footnote{http://archive.eso.org}.

The archival near-infrared NACO/VLT observations of Haro~6-10~N and S, obtained in December 2003 (Program ID:~072.C-0321(A), PI:~J.Bouvier), have been also included in our analysis. These data are also available from the ESO/ST-ECF Science Archive Facility$^1$. While the HST images are unpublished, the binary parameters derived from the NACO observations were already published, but  only in the context of an high-resolution survey of multiple young stellar systems \citep[][]{Ducheneetal2007}.

\subsection{Mid-infrared spectroscopic and interferometric observations}
\label{TIMMI2obs}
Haro~6-10 was spectroscopically observed in the mid-infrared with the TIMMI2 instrument mounted at the 3.6\,m ESO telescope at La Silla (Chile) during two campaigns in February and December 2002, see Table~\ref{timmi2}. The TIMMI2 pixel-scale in the N-band was 0.45\arcsec. The slit width used during the observations was 3.0\arcsec and 1.2\arcsec in February and December, respectively.

\begin{figure}
\centering
\includegraphics[width=9cm]{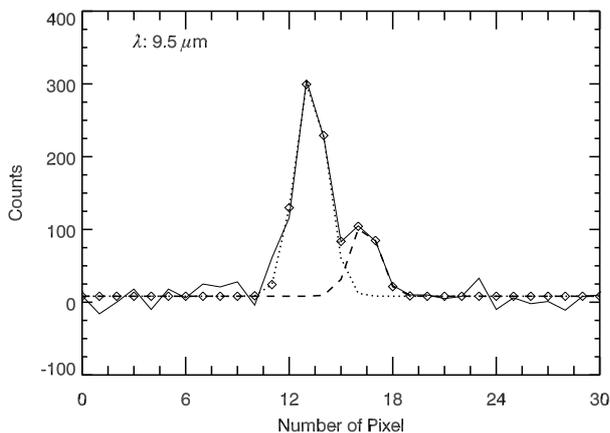}
   \caption{Perpendicular (to the dispersion direction) cut of a TIMMI~2
     bi-dimensional spectrum obtained on December 26 along with the double
     Gaussian function fit utilised for the spectra extraction. Data are shown as
     solid line, fit with diamonds. The fit to each component is shown as
     dotted line for the northern component and dashed line for the southern
     component. The TIMMI2 pixel-scale in the N-band was 0.45\arcsec.}
     \label{fitspec}
\end{figure}

We obtained two mid-infrared interferometric measurements of both Haro~6-10~S and Haro~6-10~N with MIDI at the VLTI during the Guaranteed Time Observations (GTO) in December 2004 and November 2005. The first observations were carried out using the baseline UT2-UT3, and the second observations with the baseline UT3-UT4. Details of the observations are listed in Table~\ref{obs_gvtau}. The MIDI slit width was 0.5\arcsec. The interferometric observations with MIDI were performed using the standard procedure for the high sensitivity mode which consists of 1.) acquisition images taken in the lower N-band around 8.7~$\mu$m, 2.) measurements of the spectrally dispersed correlated flux, 3.) single-dish spectra from each telescope. Spectrally resolved visibilities are then derived as the ratio of the correlated flux and the single-dish spectra. The spectral resolution obtained in the observations of Haro~6-10 is $\lambda/\Delta\lambda\sim$30 as the prism was used as dispersive element. 
A detailed description of the MIDI observing sequence is reported, e.g., in \citet{Ratzkaetal2009}.

Soon after each scientific observation, the sequence is repeated for a standard star of known diameter. These calibrators are required to determine instrumental and atmospheric effects. The calibrators, partly spectro-photometric, also served to calibrate the absolute flux as well as the correlated fluxes. We checked the stability of the transfer function using all the calibrators observed during each night.  The final calibration has been done using the calibrators closer in time to the science target (see Table~\ref{obs_gvtau}).

\section{Data reduction and results}
\label{datared}
\subsection{Optical HST photometry}
We used the calibrated HST data of Haro~6-10 already processed by the standard data reduction pipeline\footnote{http://archive.eso.org/archive/hst/}, which are available at the ESO/ST-ECF Science Archive Facility. Cosmic rays in each single exposure have been removed using a specific IDL code \citep{vanDokkum2001}. Haro~6-10~S is saturated in the long exposure of 600~sec in the F606W filter. Figure~\ref{images} shows a stack of the F606W 200s and 600s exposures (with Haro~6-10~S replaced by the non-saturated pixel in the 200~sec exposure), and the 120~sec exposure obtained with the F814W filter.

Aperture photometry of the short exposure in both filters (F606W and F814W) was carried out on Haro~6-10~N and Haro~6-10~S. An aperture of 0.50\arcsec (corresponding to 5 pixels in the F606W filter, 10 pixels in the F814W filter, respectively) has been adopted for the aperture photometry of the bright object Haro~6-10~S. The sky was estimated in the two filters in an annulus between 0.55\arcsec and 1.55\arcsec and 0.6\arcsec and 1.2\arcsec, respectively. An aperture of 0.2\arcsec has been adopted for the aperture photometry of the faint infrared companion Haro~6-10~N in both filters. Due to a non-perfect cleaning of the cosmic rays and due to saturation of the central pixel, the long 600~sec exposure in the F606W filter cannot be used for aperture photometry.

The final magnitudes have been computed, including the gain ratio correction of the PC1 and WF3 cameras and using the zero points and gain ratios from
\citet{Holtzmanetal1995}. The values obtained are listed in Table~\ref{NACOphot}.

\begin{table}
\caption{Calibrated magnitudes obtained with HST and VLT/NACO.}
\label{NACOphot}
\centering
\begin{tabular}{crrr}
\noalign{\smallskip}
\hline
\hline
\noalign{\smallskip}
&Haro~6-10~N & Haro~6-10~S&flux \\
&[mag]&[mag]&ratio\\
\noalign{\smallskip}
\hline
\noalign{\smallskip}
V (F606W)   & 23.445$\pm$0.084 & 17.169$\pm$0.003 & 0.003\\
I (F814W)   & 21.041$\pm$0.059 & 15.012$\pm$0.002 & 0.004\\
H           & 14.2  $\pm$0.1   & 11.1  $\pm$0.1   & 0.06\\
Ks          & 10.7  $\pm$0.1   &  9.2  $\pm$0.1   & 0.24\\
\noalign{\smallskip}
\hline
\end{tabular}
\end{table}

\subsection{Near-infrared photometry with NACO.}
NACO observations of Haro 6-10 were reduced with the data reduction pipeline provided by ESO\footnote{http://www.eso.org/sci/data-processing/software/pipelines/naco/}. Both components of the binary system were saturated in the L-band exposure. We derived the zero-points of the photometry from standard stars observed at the beginning and the end of the night. An aperture of 0.2\arcsec (corresponding to 15 pixels) has been adopted for the aperture photometry and the sky was estimated in an annulus between 0.30\arcsec and 0.35\arcsec. The results of the aperture photometry in the H- and Ks-bands are reported in Table~\ref{NACOphot}.

\subsection{TIMMI2 mid-infrared spectroscopy}
\label{timmi2sp}

The northern, infrared component brightens at 13\,$\mu$m from about 20\,Jy in February to about 30\,Jy in December 2002. The southern component tends to become fainter at short wavelengths with a less pronounced or maybe broader silicate absorption feature.
The derived spectra show significant temporal changes in the overall shape and absolute flux levels. 
 Individual spectra of both binary components have been already presented in \citet{przygodda04}. 
The spectra of Haro 6-10~S and its infrared companion, Haro 6-10~N, are only separated by a few pixels on the TIMMI~2 detector: For this reason we reanalysed the data of February 6 and December 26 with a different method to check how large the effect of the extraction method on the spectral variations is. 
With the aim of optimising the spectrum extraction we have first fitted the preprocessed bi-dimensional TIMMI~2 spectrum with a double Gaussian for each wavelength column (Figure~\ref{fitspec}). In this way the two spectra of each binary component are traced. Besides the fitted spectra, the fitting routine delivers synthetic images that can be compared to the original files. We found that, except the wings of the lines, the results of the fits are of good quality. The calibrators were also fitted and extracted with a Gaussian function to avoid any systematic error.

\begin{figure}
\centering
\includegraphics[width=9cm]{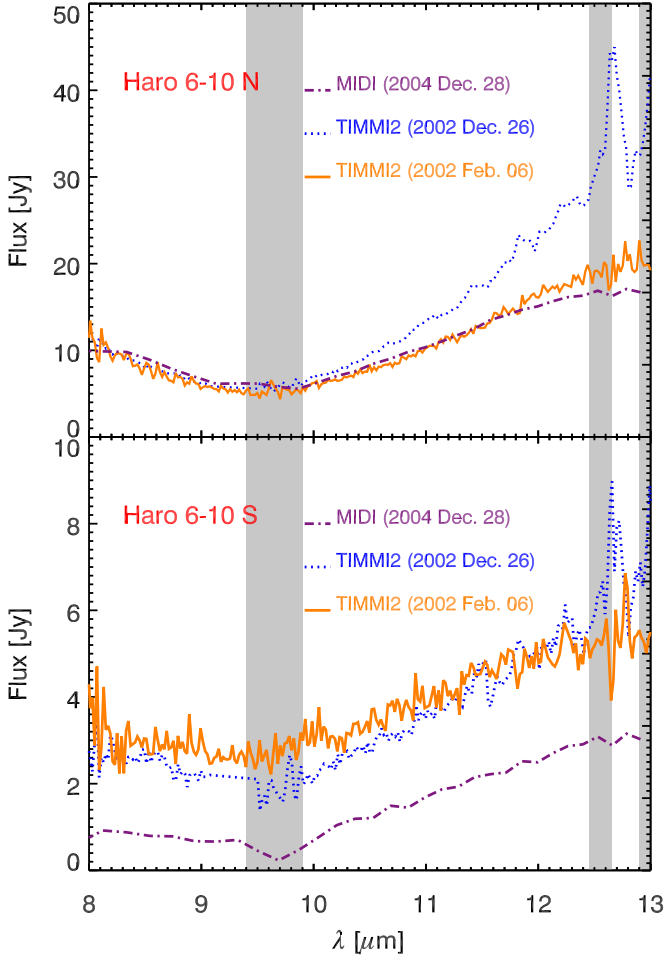}
   \caption{TIMMI~2 spectrum of Haro~6-10 N (IRC) (top) and Haro~6-10 S
     (below). The grey highlighted pattern correspond to strong atmospheric
     O$_3$ and CO$_2$ absorption bands at 9.6 and 12.5~$\mu$m, respectively. The measurements in these regions are not
     considered in our following analysis.}
      \label{timmi2spectra}
\end{figure}

The correction for telluric absorption as well as the absolute flux calibration was achieved by means of observations of spectro-photometric standard stars observed at different airmasses during each night; the extracted spectrum of each standard star was flux-calibrated using the spectral templates by \citet{Cohenetal1999}. Finally the spectra of Haro 6-10~S/N were scaled with the flux calibrated standard star spectra. The wavelength dispersion solution derived by \citet{przygodda04} was applied. The final TIMMI~2 spectra are shown in Fig.~\ref{timmi2spectra}. 


The spectra of the southern source are found to be stable in the course of 2002. 
This ensures the reliability of the new extraction method used. 
The TIMMI2 spectra show a similar shape than the MIDI spectrum taken two years later (details in section~\ref{md}), but the absolute level is about three times larger. The TIMMI2 spectra of the northern component show changes in flux in the upper part of the N-band \citep[also shown by ][]{przygodda04}. However, the MIDI spectrum of Haro 6-10~N resembles the TIMMI2 data from February 2002 very well. The instrumental FWHM increases from 0.46\arcsec at 8~$\mu$m to 0.74\arcsec  at 13~$\mu$m. The effective FWHM, measured directly from the spectrum is 0.9\arcsec and does not depend on the wavelength. We notice that the differences between the TIMMI2 and MIDI spectra concern only the southern component of the binary system, Haro~6-10~S, which is the fainter one. Although the different FWHM of the two instruments, TIMMI2 and MIDI, might introduce in the TIMMI2 spectra background emission from the surrounding nebulosity, this would have affected both the spectra of Haro 6-10~N and S; however, since the northern component is about 10 times brighter than the southern one, the contamination from the background  might significantly increase only the flux of the fainter component Haro~6-10~S. Another problem might be that the deblending of the two TIMMI2 spectra is is highly contaminated by the northern brighter source. However, the latter can probably not alone explain the large discrepancy between the MIDI and the TIMMI2 spectrum of Haro 6-10~S.

\subsection{MIDI Data} 
\label{md}
The data reduction of interferometric data with MIDI has been performed with the MIA+EWS~1.5 data reduction software provided by the MIDI consortium\footnote{http://www.strw.leidenuniv.nl/$\sim$koehler/MIA+EWS-Manual/index.html}. The results of the data reduction are the raw correlated spectrum F$_{\rm raw}^{\rm corr}$ and the total (single-dish) raw spectrum F$_{\rm raw}$. We here use the results of the EWS package, which determines the correlated flux F$_{\rm raw}^{\rm corr}$ based on a ``coherent'' algorithm. MIA, based on power spectrum analysis, was used to confirm the results.

\subsubsection{Visibility}

We compute the raw visibilities from 
\begin{equation} \label{VR}
V_{\rm raw}=\frac{F_{\rm raw}^{\rm corr}}{F_{\rm raw}},
\end{equation}
where F$_{\rm raw}$ is the geometrical mean of the total fluxes of the incoming beams from telescope A and B, i.e., $F_{\rm raw}=\sqrt{F_{\rm raw}^{\rm A}\cdot F_{\rm raw}^{\rm B}}$.

The calibrated visibilities $V_{\rm cal}$ of the science object are obtained by dividing the raw visibilities, $V_{\rm raw}$, by the instrumental visibilities derived from one or more calibrators observed within the same night, usually the closest in time. In Table~\ref{obs_gvtau} we compile the calibrators observed during each night,  marking those, observed closest in time to the science target, that have been used to calibrate the final visibilities. The other calibrators were used to check the stability of the transfer function. The error associated to the calibrated visibility of the science targets is the standard deviation resulting from the calibration with different calibrator stars for each night.



\subsubsection{Spectroscopy}

As part of the observing scheme, MIDI acquires the N-band spectrum of the target (spectral resolution equal to that of the interferometric signal) with both telescopes separately. MIDI spectra were extracted using the MIA+EWS software. 
In this regards we used the three observations of the spectro-photometric standard star HD~37160 obtained in December 2004 (see Table~\ref{obs_gvtau}). In Fig.~\ref{timmi2spectra} we compare the MIDI and TIMMI2 spectroscopy. The results are very similar for Haro~6-10~N, while for Haro~6-10~S TIMMI2 yields a substantially higher absolute flux. This is likely due to the strong effect of the deblending on the fainter component and/or to the higher spatial resolution of the VLT compared to the 3.6m telescope.
The TIMMI2 spectrum of Haro~6-10~S might be contaminated by the northern component (see also discussion in Section~\ref{timmi2sp}) or by large-scale background emission. 

\subsubsection{Correlated flux}

To flux calibrate the correlated spectrum one approach is to use the following equation
\begin{equation} \label{VC}
F_{\rm cal}^{\rm corr}=V_{\rm cal}\cdot F_{\rm cal},
\end{equation}
where $V_{\rm cal}$ is the calibrated visibility and $F_{\rm cal}$ is the flux-calibrated spectrum. An alternative approach is to directly calibrate the raw correlated flux, F$_{\rm raw}^{\rm corr}$, by means of spectro-photometric standards. The results obtained using the two methods for bright sources agree well, while there are some discrepancies for the faint source Haro~6-10~S. The correlated fluxes for our analysis are all consistently computed with Equation~(\ref{VC}).

\section{Analysis}
\label{analysis}
In this section, we present the analysis of our multi-wavelength observations of Haro~6-10. We analyse first the large-scale structure of the binary by investigating visual and near-infrared images as well as mid-infrared spectra. In a second step, we examine the geometrical properties of the protoplanetary discs around the binary components through modelling MIDI mid-infrared visibilities and correlated fluxes.   

\subsection{Large-scale morphology in the optical and near infrared}
\label{optical-nir}
\begin{figure}
\centering
\includegraphics[width=5cm]{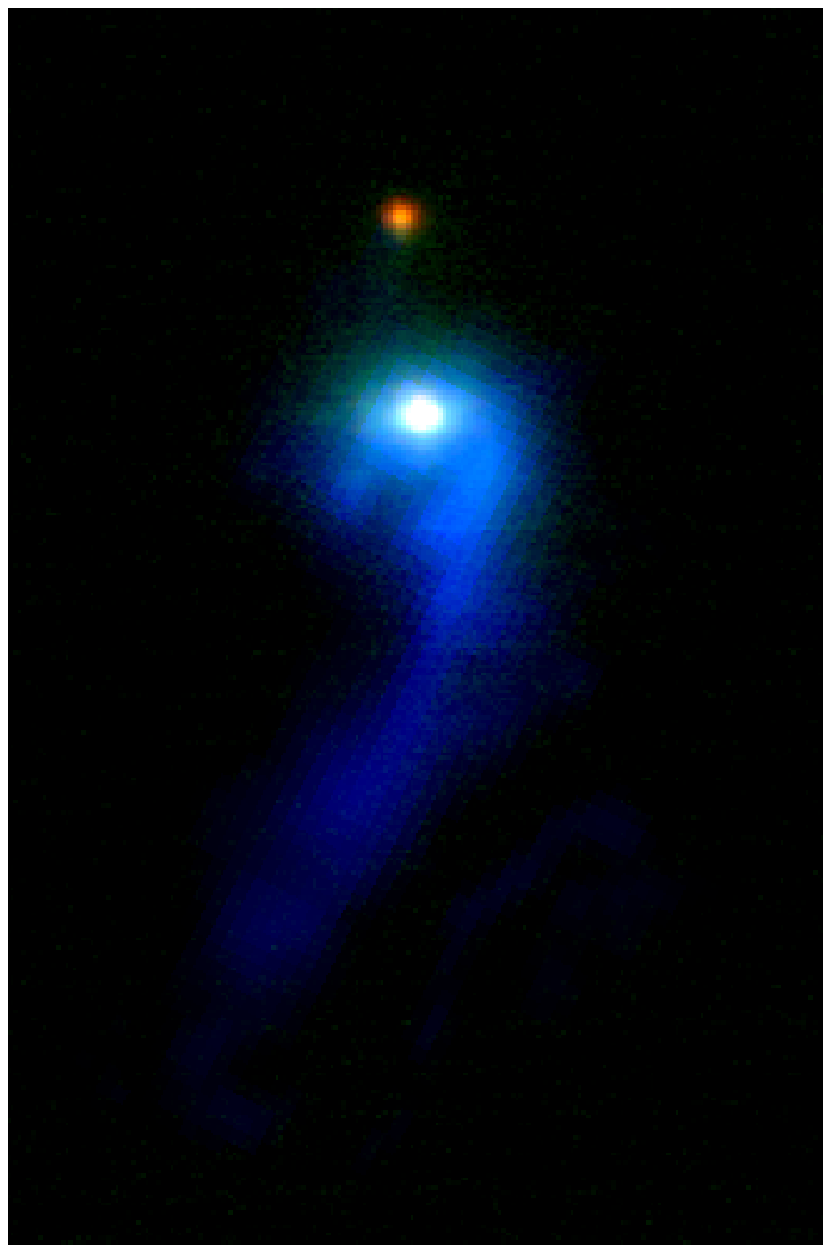}
\includegraphics[width=9cm]{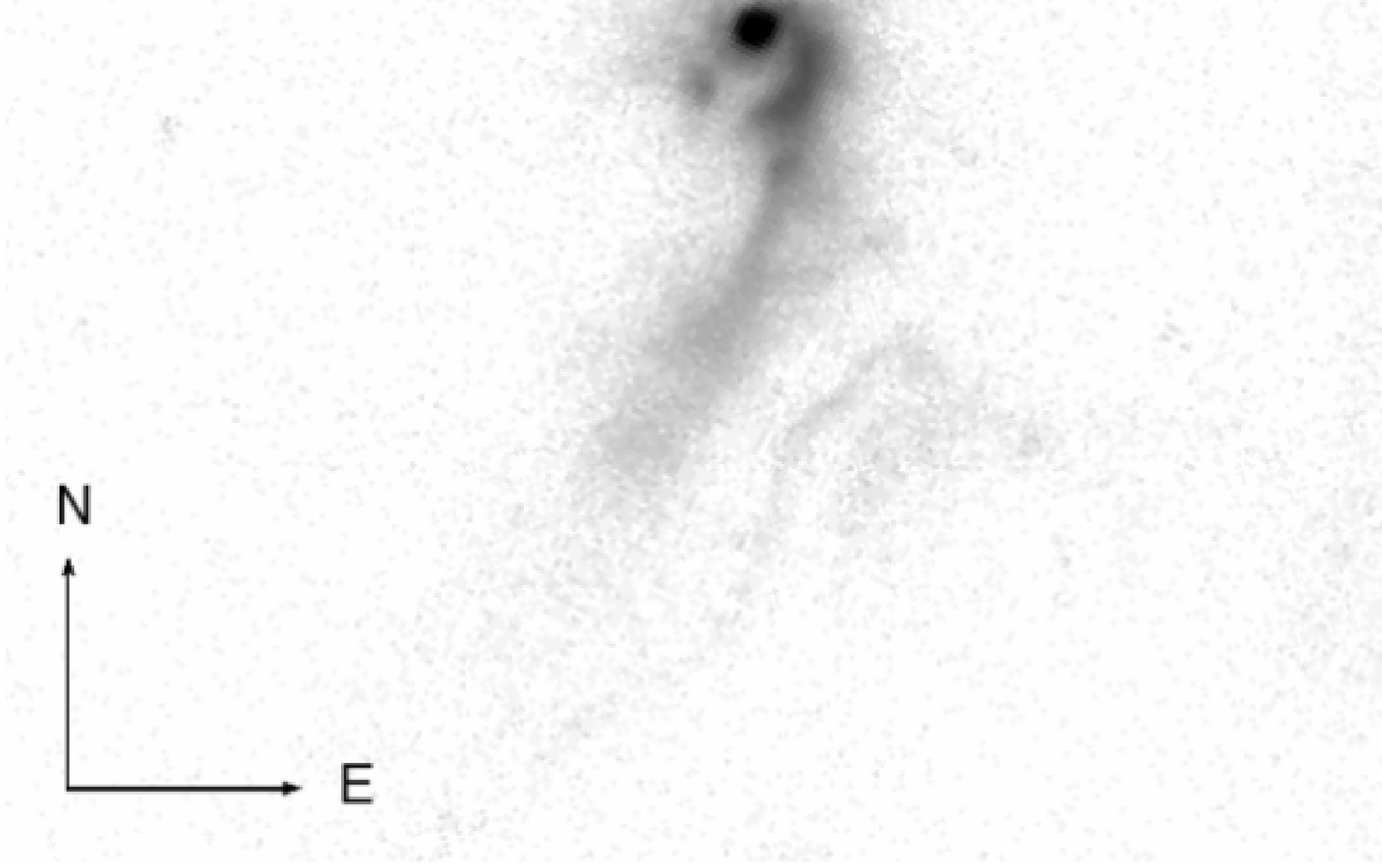}
   \caption{{\it Upper panels:} Color-composite image of Haro~6-10 in logarithmic scale. The blue, green, and red colours correspond to the photometric filter F606W (HST) and the H and K$_s$ band (NACO). North is up, East is right.  {\it Lower panels: } HST images obtained  using the F606W ($\sim$V~band) and F814W($\sim$I~band) filters and NACO images in the H, K$_s$, and L'-band. 
     }
      \label{images}
\end{figure}

Figure~\ref{images} shows the WFPC2/HST optical images and the VLT/NACO near-infrared images. Both components of the binary system are resolved from the optical to the mid-infrared wavelengths. While in the optical the infrared companion is only barely detected, it becomes brighter in the near-infrared. The flux of Haro~6-10~N is 0.3\% of the flux of Haro~6-10~S in the visual and increases up to 24\% in the K band. The photometric results and the flux ratios are listed in Table~\ref{NACOphot}. In the N-band Haro~6-10~N is brighter than the southern companion (Figure~\ref{timmi2spectra}).

The resolution of 0.05\arcsec/px and 0.1\arcsec/px of the HST images allows to disentangle structures and the filaments in the environment of the Haro~6-10 binary system. An arc-like structure is clearly detected up to a distance of $\sim$6\arcsec S and $\sim$3\arcsec W from the southern component, where it fades, together with other north-south oriented filamentary structures, in the underlying nebulosity. Previous observations \citep[e.g.,][]{Leinertetal2001, Koreskoetal1999} discovered the short and bright arc-like structure east of Haro~6-10~S (see Section~\ref{previous}). Another prominent structure is a bright knot less than 1\arcsec south-west of Haro~6-10~S. In the H- and K-band these filament structures become fainter and are undetectable at longer wavelengths. The southern component appears to be barely resolved in the E-W direction.

\begin{figure*}
\centering
\includegraphics[width=13cm]{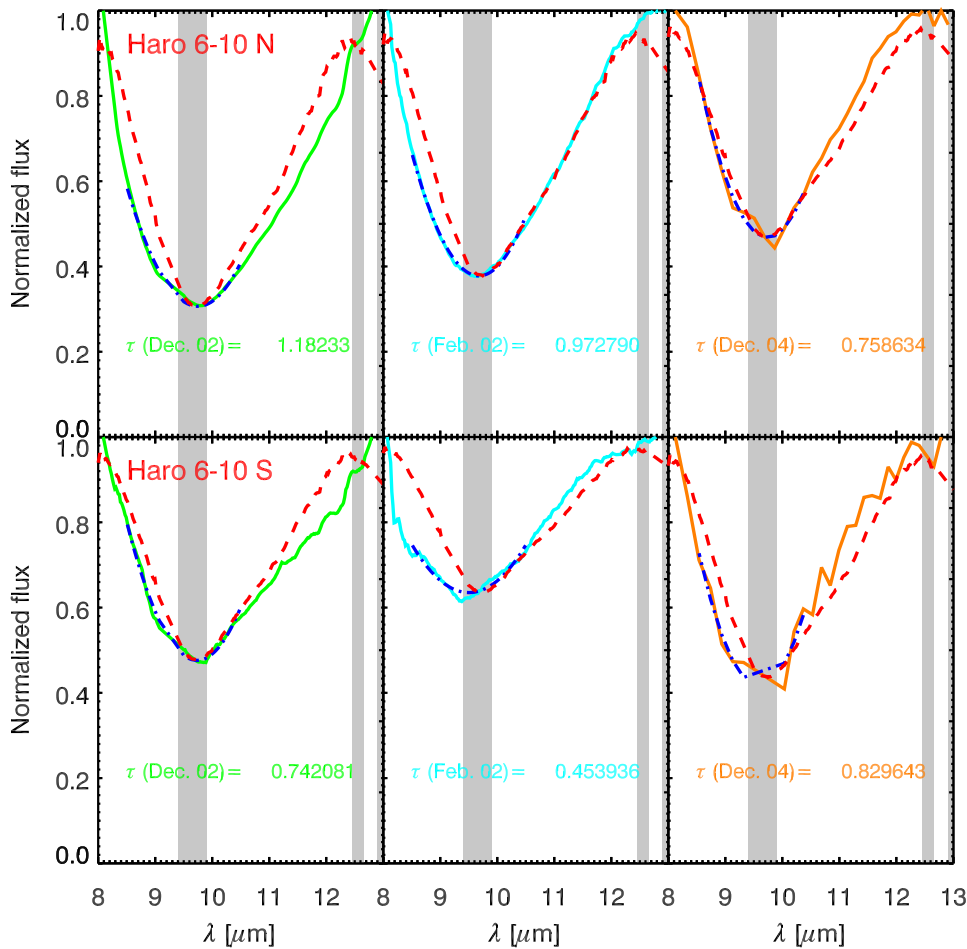}
   \caption{Normalised spectra of Haro~6-10~N (top) and Haro~6-10~S (bottom)
     obtained with TIMMI2 on February and December 2002 and with MIDI on
     December 2004, with the computed maximum optical depth. A second-order
     polynomial has been fitted to the data between 8.8 and 10.2~$\mu$m
     (dot-dashed lines) to compute the maximum optical depth ($\tau$). For
     comparison the ISM silicate profile from \cite{Kemperetal2004} is
     overplotted (dashed red lines).}
      \label{tau}
\end{figure*}

\subsection{Optical extinction and dust composition}
\label{anTIMMI2}
To derive the optical depth of the circumstellar material we fitted the continuum with a first-order polynomial to the TIMMI2 spectrum between 8 and 13~$\mu$m, assuming that at these wavelengths only the continuum contributes to the observed spectrum. The depth of the absorption feature was then determined with a second-order of polynomial fit. 
A similar approach was presented by \cite{Quanzetal2007}. We compute the optical depth $\tau_{\rm silicate}$ using the following equation:
\begin{equation}
F_{\rm obs}(\lambda)=F_{\rm cont}(\lambda)\,\cdot \,e^{-\tau_{\rm silicate}}{\rm\ .}
\end{equation}
And from this we derive the extinction $A_{\rm silicate}$ by
\begin{equation}
A_{\rm silicate}= -2.5 \,\cdot \,log_{10}(e^{-\tau_{\rm silicate}}){\rm\ .}
\end{equation}
The optical extinction $A_{V}$ was computed using the relation $A_{V}/\tau_{9.7}=19.3$ from \cite{Mathisetal1998}. All the results are listed in Table~\ref{tau_table}. We derive an average visual extinction of 18.7$\pm$4.1~mag for Haro~6-10~N and 13.0$\pm$3.8~mag for Haro~6-10~S. Our values are higher but with a similar ratio than the values obtained by \cite{Leinertetal2001}. From the ice-absorption feature they derived 10.6$\pm$0.8~mag for Haro~6-10~N and 7.0$\pm$0.7 mag for Haro~6-10~S. For such a young system, this absorption is most probably caused by a circumbinary envelope. The spectra of Haro~6-10~N and Haro~6-10~S are compared in Fig.~\ref{tau} with the profile of the silicate dust present in the interstellar medium (ISM). This profile was measured by \cite{Kemperetal2004} and is shown as dashed line. The spectra of the binary components are very similar to the ISM profile. However, the ISM absorption looks broader compared to the absorption observed towards the binary. The quality of the spectra does not allow to conclude whether these deviations are due to  a variation in the chemical composition or on the grain size of the dust around the binary system.
A further analysis of the dust composition using, e.g., a simple fitting of the N band spectrum \citep[as done for Herbig and TTauri stars, e.g., by][]{Bouwmanetal2001, Schegereretal2006} is beyond the goal of this paper. 
\subsection{Small-scale morphology: Geometry of the discs}
In this section we introduce the modelling strategy adopted to interpret the MIDI interferometric measurements. 
The characteristic size of the emitting region was firstly computed assuming a Gaussian brightness distribution. In a second step, the emitting sizes are computed fitting the two observations available per object with a {\it Two Black-Body model} with a Gaussian distribution of the emitters.
An alternative approach would be to do the simultaneous modelling of the spectral energy distribution (SED) and the mid-infrared interferometric measurements, using a radiative transfer model \citep[e.g.,][]{Schegereretal2008, Ratzkaetal2009, Schegereretal2009}. This has the advantage to reproduce in a more realistic way the physical structure of the circumstellar disc. However, the degeneracy in the parameter space left when using only a few visibility measurements can be reduced only by taking supplementing information into account, e.g.,  high-resolution images at different wavelengths \citep[e.g.,][]{Wolfetal2003,Wolfetal2008,Sauteretal2009}. 

\subsubsection{Gaussian brightness distribution}
\label{gbd}
To estimate the size of the emitting region from the observed visibility on the basis of a Gaussian brightness distribution is a reasonable first approximation already applied for barely resolved (high visibility) objects \citep[e.g., by ][]{Quanzetal2006}. The visibility of such a distribution is computed as   
\begin{equation} 
V(f) =  \exp\left(-3.56~f^2~\Theta^2\right){\rm \ ,}\label{eq5}
\end{equation}
where $\Theta$ is the Full Width Half Maximum (FWHM) of the Gaussian distribution in arcsec, and $f$ the spatial frequency in arcsec$^{-1}$. For each baseline the emitting sizes of the two discs (using the FWHM) are reported in Table~\ref{Gaussian} as function of wavelength. These values are given in astronomical units (AU) assuming a distance of 140 pc. We average the results obtained within 3 wavelengths bins: 9.8 - 7.9~$\mu$m, 12.0 - 10.0~$\mu$m and 13.0 - 12.1~$\mu$m. The associated errors are the standard deviations. We note that the emitting size increases with wavelength. Such a behaviour is expected for thermal disc emission, where the temperature of the disc decreases with the distance from the central star.
This implies that we are resolving two extended structures around the central star. Despite the large appearance 
discrepancy of the two sources in the visible, the derived sizes of the two dusty components in the mid-IR are very similar, even when taking into account the uncertainties on the position angle of 
each disk. The three wavelengths bins reflect the three regimes of opacity: the `hot' continuum, the increase of size 
due to the silicate opacity, and then the `cool' more extended continuum near 13 micron.

      

\begin{table}
\caption{Optical depths and extinction values as derived from the spectra in Fig.~\ref{tau}. }
\label{tau_table}
\centering
\begin{tabular}{crrrrr}
\hline
\hline
\noalign{\smallskip}
Object      & Date          & $\tau_{\rm silicate}$ & $\lambda_{\rm silicate}$ & $A_{\rm silicate}$ &$A_{V}$ \\
\noalign{\smallskip}
\hline
\noalign{\smallskip}
Haro~6-10~N  & 2002 Feb.     &		1.0	& 9.7 & 1.1 & 18.8\\
	                 & 2002 Dec.     &		1.2	& 9.7 & 1.3 & 22.8\\
	                 & 2004 Dec.     &		0.8	& 9.7 & 0.8 & 14.6\\
Haro~6-10~S & 2002 Feb.      &		0.5	& 9.5 & 0.5 &   8.8\\
	                 & 2002 Dec.     &		0.7	& 9.7 & 0.8 & 14.3\\
	                 & 2004 Dec.     &		0.8	& 9.3 & 0.9 & 16.0\\
\noalign{\smallskip}
\hline
\end{tabular}
\end{table}

\begin{table}
  \centering
  \caption{Derived emitting size in [AU] using the FWHM of a Gaussian brightness distribution.}
  \label{Gaussian}
  \begin{tabular}{lllll}
    \hline
    \hline
    \noalign{\smallskip}
    \multicolumn{1}{c}{}& 
    \multicolumn{2}{c}{Haro 6-10~N}&
    \multicolumn{2}{c}{Haro 6-10~S}\\ 
     \noalign{\smallskip}
    \multicolumn{1}{c}{$\lambda$} & 
     \multicolumn{1}{c}{UT2-UT3} & 
     \multicolumn{1}{c}{UT3-UT4} & 
     \multicolumn{1}{c}{UT2-UT3} & 
     \multicolumn{1}{c}{UT3-UT4}	 \\
     \noalign{\smallskip}
     \multicolumn{1}{c}{[$\mu$m]} & 
     \multicolumn{1}{c}{[28 m]} & 
     \multicolumn{1}{c}{[62 m]} & 
     \multicolumn{1}{c}{[38 m]} & 
     \multicolumn{1}{c}{[58 m]} 	\\
   \noalign{\smallskip}
    \hline
    \noalign{\smallskip}
     7.9 -- 9.8  & 3.4$\pm$0.6 & 2.5$\pm$0.3 & 1.8$\pm$0.4 & 1.0$\pm$0.2 \\
    10.0 -- 12.0 & 4.3$\pm$0.1 & 3.2$\pm$0.1 & 2.9$\pm$0.2 & 1.4$\pm$0.3 \\
    12.1 -- 13.0 & 4.6$\pm$0.8 & 3.5$\pm$0.2 & 3.2$\pm$0.6 & 2.2$\pm$0.3 \\
    \noalign{\smallskip}
    \hline
  \end{tabular}
\end{table}

\begin{figure}
\centering
\includegraphics[width=9cm]{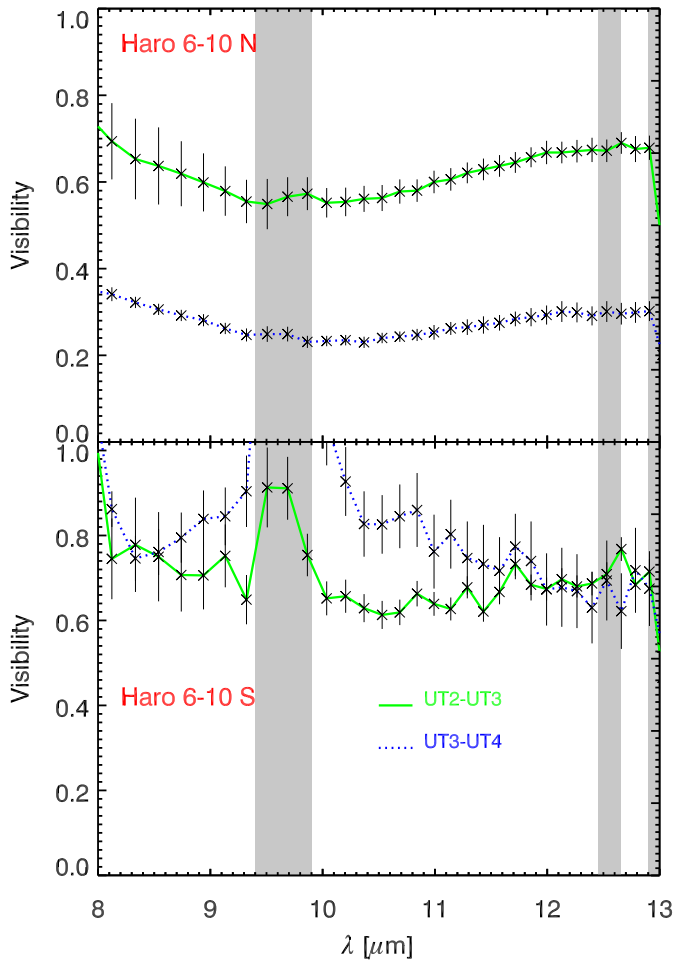}
   \caption{Calibrated visibilities of Haro~6-10~N (top) and
     Haro~6-10~S (bottom). 
     }
     \label{vis}
\end{figure}


\subsubsection{{\it The `Two Black Body' Model}}
\label{2bbm}
We note that the change in the slope of the visibility with wavelength may suggest the presence of two different emitting regions
contributing to the mid-infrared flux (see Figure~\ref{vis}). 
%
We introduce here a more complex (geometrical) model which takes into account also the spectrum and correlated flux of the emitting source. 
We model simultaneously the MIDI spectrum (integrated over the beam width of 0.5\arcsec) and the correlated fluxes corresponding to spatial scales at 10~$\mu$m of 0.07\arcsec\, and 0.03\arcsec\, for the 28~m and 62~m projected baselines, respectively.

In a circumstellar disc, in first approximation, we expect a radial decrease of the temperature as a function of the distance from the central star. Taking into account the vertical extension of the disc, a decrease of the disc temperature is expected also from the surface to the mid-plane. The mid-infrared flux is therefore expected to come from the inner part of the disc close to the central star. In this simplified view we approximate the emitting source with two black-bodies of different temperatures: each black-body is characterised by a Gaussian brightness distribution \citep[a similar approach was used  by][]{Jaffeetal2004}. In order to reproduce the dust absorption feature between 8 and 13~$\mu$m the silicate absorption profile is included. In particular, we use the absorption profile derived from ISO\footnote{Infrared Satellite Observatory} towards the centre of the galaxy \citep{Kemperetal2004}, which is consistent with the absorption profile of Haro~6-10~N and S (see Section~\ref{anTIMMI2}). The details of the model used are presented in Appendix~\ref{models}.

We used position angles of the discs (Haro 6-10~N: 140 $^o$, Haro 6-10~S: 120$^o$) as derived from the two different jets originating from each source  \citep[][T. Beck, private communication]{Becketal2010}. This has been done to ensure that the number of free parameters is not biasing the results.

\begin{figure*}
    \includegraphics[width=9cm]{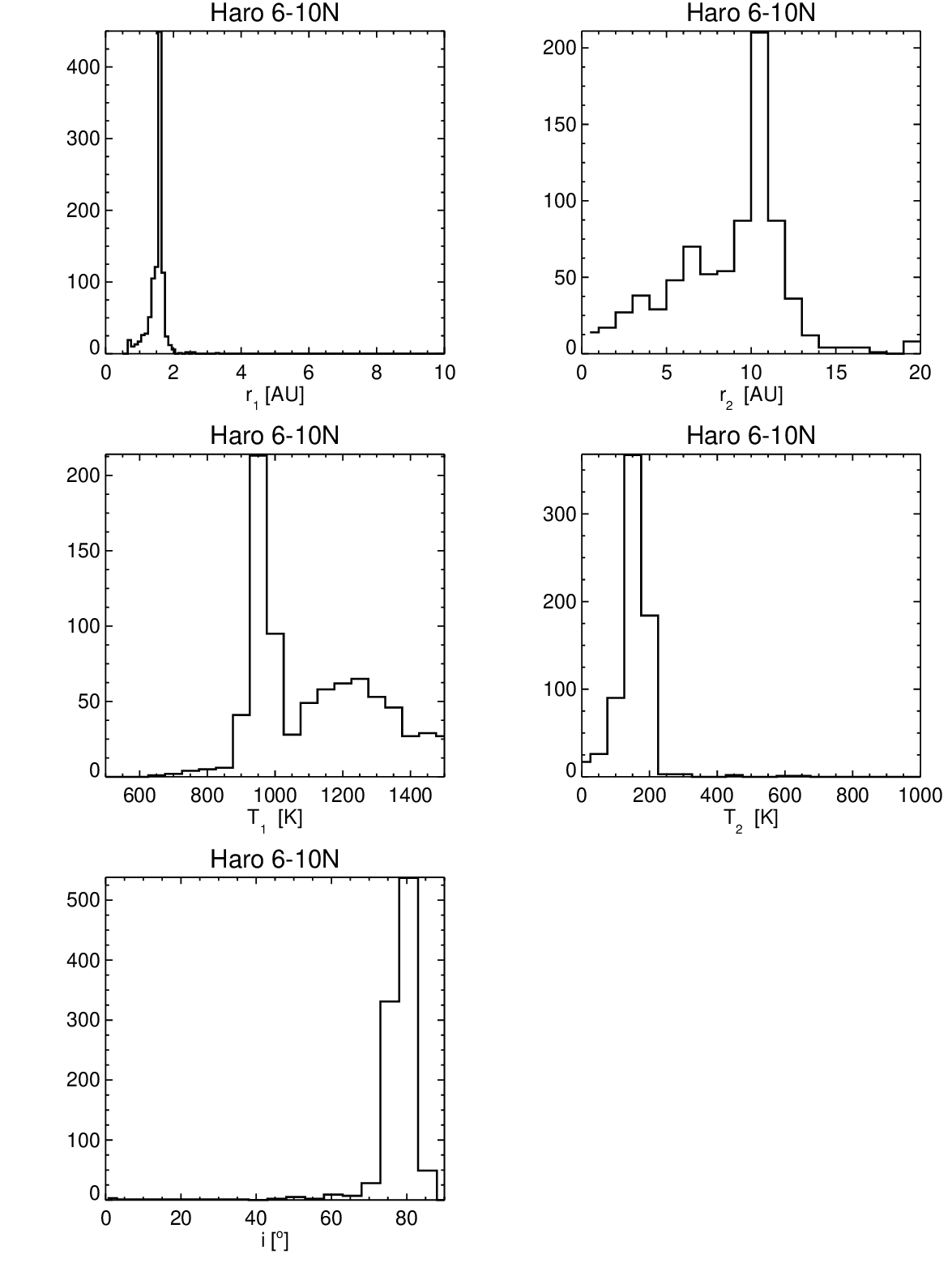}
    \includegraphics[width=9cm]{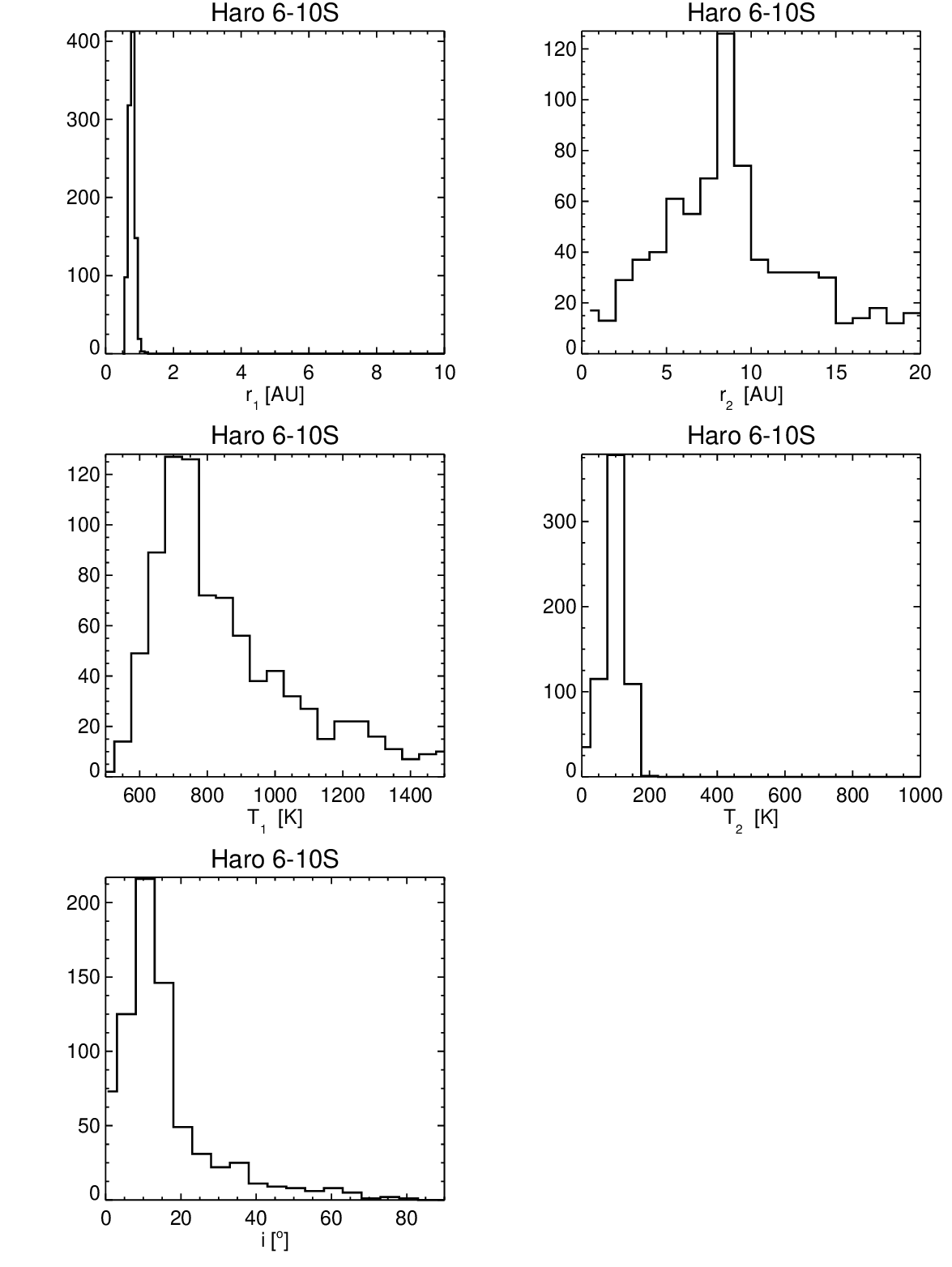}
    \caption{Histograms of the best fit parameters computed using 1000 Monte
      Carlo simulations for Haro~6-10~N (left) and Haro~6-10~S (right).}
    \label{figMC_BBG_N}
  \end{figure*}

A first {\it best} model was found minimising the reduced $\chi^2$ ($\chi^2$
  divided by the number of free parameters of the fit). 
 In order to have a more reliable result, following the approach presented by  \citet{Fedeleetal2008}, 
  \begin{enumerate}
\item we created 1000 random new data-sets of measured visibilities around the observed visibilities, assuming a normal distribution of the visibility errors;
\item we computed the best-fit model for each new data-set. 
  \end{enumerate}
The final best fit parameters and associated uncertainties 
  correspond to the mean and the standard deviation of the 1000 fits. 
Fig.~\ref{figMC_BBG_N} shows the histograms of the best fit parameters computed for each fit, and the final best fit values are listed in Table~\ref{tabbestfit}.\\
For Haro~6-10~N we found that the emission within 1.5~AU from the central star has a temperature of 900$K$ and a second colder emission at 150$K$ originates within 10~AU from the central star. For Haro~6-10~S we found an emission within 1.0~AU from the central star with a temperature of 900$K$ and a second colder emission at 100$K$, originating within 7~AU  from the central star. The disc of Haro~6-10~N is seen close to edge-on ($i = 80^o$), while Haro~6-10~S is almost seen  
face-on ($i = 10^o$). 
  The derived inclinations of the binary components are consistent with the fact that the central star of Haro 6-10 S is visible in the optical, while the star of Haro 6-10~N is obscured by an additional extinction represented by the circumstellar disc and is only barely visible in the optical.  
These results are also consistent with the general appearance of the binary system described in Section~\ref{optical-nir}.

  In the case of Haro~6-10~N, the model fits the observations very well, 
  while for Haro~6-10~S the model fitting is less accurate; 
  the model reproduces the mid-infrared spectrum,
  the correlated  spectra, and the absolute value of the visibilities, but does not
  fit their shape. 
We thus additionally tested the robustness of our results to be sure that the Monte Carlo simulations did not only find a local minimum in the parameter space instead of the absolute minimum. Since the best defined parameters are the disc inclinations, we changed their initial values, for Haro~6-10~N to a face-on disc ($i = 3^o$) and for  Haro~6-10~S to an edge-on disc ($i = 80^o$). Also in this case the Monte Carlo simulations converge to the results previously obtained. We also run the whole modelling with the position angles as additional free parameters. Although the derived position angles differ from the observed ones by $\sim70^o$ for Haro 6-10 N and  $\sim50^o$ for Haro 6-10~S, the results for the remaining parameters are almost identical to the above given values.

\begin{figure}
\includegraphics[width=8.3cm]{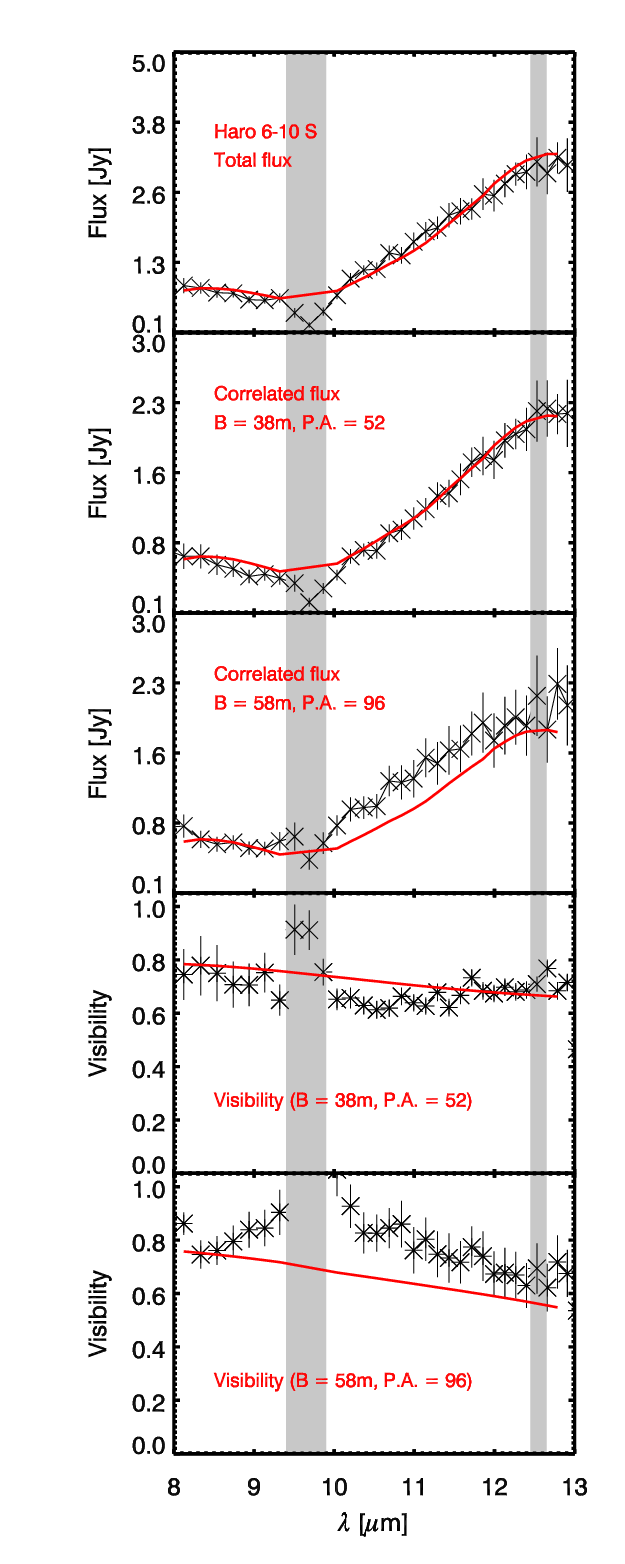}
   \caption{MIDI spectrum, correlated fluxes and visibilities for Haro~6-10~S.
     The red solid line represents the model
     best fit.}
      \label{BBG_S}
\end{figure}

Given the higher uncertainties of the measurements for the southern component and the poorer fits (Fig.~\ref{BBG_S}), the formal errors derived for its disc parameters are most probably a lower limit. The simple size estimates derived in Section~\ref{gbd} are also consistent with higher inclinations (Table~\ref{Gaussian}). We further want to emphasise, that our approach strongly relies on the assumption that the components of Haro~6-10 can be described by a superposition of two Gaussian brightness distributions. If the true brightness distribution would show significant deviations, our results would be biased. Also the interpretation of the value found for the inclination (defined as the elongation of the Gaussians) then would be no longer straightforward. Unfortunately, such deviations can not be ruled out with the limited number of measurements at hand. However, Gaussian brightness distributions have proven to be a reasonable assumption for a wide variety of centrally heated objects.

\begin{table}
\caption{Best fit parameters of the `two black body' model.}
\label{tabbestfit}
\centering
\begin{tabular}{lrr}
\hline
\hline
\noalign{\smallskip}
Parameters   & Haro~6-10~N & Haro~6-10~S \\
\noalign{\smallskip}
\hline
\noalign{\smallskip}
r$_1$ [AU]     & 1.5$\pm$0.5 	&1.0$\pm$0.5\\ 
r$_2$ [AU]     & 10$\pm$2	&7$\pm$3\\
T$_1$ [K]      & 900$\pm$100 	&900$\pm$300\\
T$_2$ [K]      & 150$\pm$50 	&100$\pm$50\\
i [$^o$]       & 80$\pm$10      &10$\pm$5\\
\noalign{\smallskip}
\hline
\end{tabular}
\end{table}

\section{Discussion}
\label{discussion}

In this section we compare our results with previous works and discuss which mechanism might have formed the binary system Haro~6-10.

\subsection{Large- and Small-Scale Structures in the Haro 6-10 Environment}
\label{previous}
The optical images of Haro~6-10 highlight some filamentary structures (as described in Section~\ref{optical-nir}) which might have the following explanations: 1) it is circumstellar material with an arc-like shape simply illuminated by the central star, or 2) it is the projection of the scattered light of a cavity in the circumbinary envelope, or 3) it represents the circumstellar material shocked by an outflow.

The existence of a cavity in an envelope has been suggested around a similar system, T~Tau~N, by \cite{Momoseetal1996} and \cite{Stapelfeldtetal1998}. In this picture, a wide-angle wind creates oblique shocks when interacting with the molecular environment. On the other hand, the existence of an outflow is supported by the resolved extended emission of H$_2$ at 6.5\arcsec in the south direction and is consistent with the filament structure \citep[][]{Doppmannetal2008}. \cite{Devineetal1999} already identified three outflow systems in their narrow-band optical images. In addition to the large-scale structure two separate outflows closer to Haro 6-10 were found. This suggests that both components are producing outflows. 
 Based on integral-field spectroscopy and polarimetry \cite{MovsessianMagakian1999} suggest the northern source as the origin of a jet with a length of 5\arcsec superimposed on the bright reflection nebula. They propose that the disc around this component is seen edge-on, while the southern component is surrounded by a less inclined disc. \cite{Stojimirovicetal2007} report on a parsec-scale Herbig Haro flow with clumpy redshifted gas pointing to the north-east. The blueshifted south-western part of the flow is contaminated by an unrelated foreground cloud. \cite{Becketal2010} found a bulk of Br$\gamma$ emission from the two sources together with an extended, fainter emission from the southern component. The same extended emission was measured also in the [Fe\,II] line and is thus clearly determining the direction of an outflow originating from this source. 

At 1.2~mm the emission was again resolved with the PdBI \citep[][]{Guilloteauetal2011} and the millimetre flux is almost the same for the two components. Under the assumption that the disc is optically thin at these wavelengths, the mass of the discs can be directly derived from the fluxes. The measured fluxes thus imply that the discs of Haro~6-10~N and S have almost the same mass. The measured large extinction associated by \cite{Stojimirovicetal2007} to a foreground cloud is also in agreement with the envelope we introduced to explain the mid-infrared absorption spectrum through both components of the binary system.
Our findings on the circumstellar discs around the two sources of the binary system are consistent with the picture drawn by the above studies. Also the misalignment of the discs and the suggested high inclination of the northern component is confirmed by our measurements. 
In particular, the almost face-on disc could explain why the emission of the spatially resolved Br$\gamma$ emission is observed blueshifted: the corresponding redshifted emission might be covered by the 
face-on disc.
Since the position angles of the discs can not be well constrained by our new data,  we did our modelling (Section~\ref{2bbm}) with fixed position angles (T. Beck, private communication) as well as with the position angles as free parameters. Both approaches lead to the same results.

\subsection{Comparison to previous work}

The main results of our study are that Haro 6-10 is embedded in a common envelope (which causes the 
mid-infrared absorption feature in both components of the binary system),
and that the two discs of the binary system are misaligned.

Observational studies of the disc inclinations in young binary systems have
been carried out in the last years, using different techniques: polarimetric
and direct observations of protoplanetary discs or indirect observations via
the position of the jet from the protostellar object, which is expected to be
perpendicular to the disc. Recently, also long-baseline infrared interferometry
was used to measure the geometrical properties of circumstellar discs in
binary systems.

Polarimetric studies on the disc orientations have been carried out 
by, e.g., \citet{Moninetal1998, Wolfetal2001, Moninetal2006}.
They all observed T Tauri stars and found that discs tend to be aligned in
young binaries. \cite{Jensenetal2004}, in particular, found a tendency for
binary system discs to be nearly (but not exactly) aligned with each other,
and for those in triple and quadruple systems to be misaligned.\\
There are also cases of misaligned discs found by observations of misaligned jets
from protostellar objects \citep[e.g.,][]{Davisetal1994}, inferred jet
precession \citep[e.g.,][]{Eisloffeletal1996, Davisetal1997}, and direct
observations \citep[e.g.,][]{Koresko1998, Stapelfeldtetal1998a}. 
A recent interferometric study of T Tau showed that the discs in this triple
system are misaligned \citep{Ratzkaetal2009}: The circumstellar disc of the
northern component is seen almost face-on, while the 
southern components are surrounded by higher inclined discs. Especially the disc around
T~Tau~Sa, the most massive star in the system and a prototypical IRC, is seen almost edge-on. The
formation process of T~Tau thus has to have been highly dynamic. 
However, these cases of misaligned discs do not represent yet a statistical
significant sample.

\subsection{How Haro~6-10 formed?}

The misalignment of the discs in the Haro~6-10 binary system might suggest the formation mechanism of the system itself. Such a misalignment cannot be caused by an infall of external material with a different angular momentum onto the binary orbit, because the accretion would have re-aligned the two discs \citep[][]{Bateetal2000}. Only the infall on one of the components might lead to a tilt of that component, when the transported angular momentum is large enough. 
Also capture of the companion is an unlikely explanation, because for 
solar-mass stars the capture rates are insufficient even when circumstellar 
discs increase the interaction cross section considerably \citep[][]{Heller1995}. For 
the central part of the Orion Nebula Cluster (ONC) with its stellar density of 
10$^4$ stars~pc$^{-3}$  \citet{Boffinetal1998} found a capture rate of only $0.1\,{\rm 
Myr}^{-1}$, i.e. the probability to form a binary star within 1\,Myr is only 
10\% . For environments like the Taurus-Auriga star-forming complex, with 
stellar densities of around 10 stars~pc$^{-3}$ the capture rate is two orders of 
magnitude lower. However, disc-assisted capture is efficient for dense 
environments and massive stars with large discs \citep[][]{MoeckelBally2006, MoeckelBally2007a, MoeckelBally2007b}. 


The most probable scenario is the formation via fragmentation of two different parts of the collapsing molecular cloud combined with other dynamical processes related to the cloud or the protostars.

However, no final conclusion about the formation scenario can be drawn yet. More about the dynamics of the system has to be known. Nevertheless, our observations of Haro~6-10 show another system that has misaligned discs. 
Many of the binaries with infrared companions might harbour an edge-on and a face-on disc.

\section{Conclusions}
\label{conclusions}
In this paper we presented a multi-wavelength study  of the young binary system Haro 6-10 combining optical images obtained with HST/WFPC2 with the diffraction-limited images obtained with VLT/NACO in the near-infrared. In the mid-infrared we combine the spectroscopic and interferometric observations obtained with TIMMI2 and VLTI/MIDI, respectively. This unique collection of data enables us  to characterise the large-scale and small-scale structures of Haro~6-10 and to draw the following conclusions:
\begin{enumerate}
\item Both components of the binary system Haro~6-10 are embedded in a common envelope. 
The envelope has a dust composition  similar to the interstellar medium.
\item Each component of the system has a circumstellar disc-like structure typical of young stars.
\item The discs are highly misaligned: the northern disc is seen almost edge-on with respect to the line-of-sight and the southern 
     component is surrounded by an almost face-on disc.
\item The most probable scenario is the formation via fragmentation of two different parts of the collapsing molecular cloud combined with other dynamical processes related to the cloud or the protostars.
\end{enumerate}
\begin{acknowledgements}
Based on observations made with the Hubble Space Telescope
    obtained from the ESO/ST-ECF Science Archive Facility. 
V.R. thanks Anna Pasquali for her help with the HST photometry and Tracy Beck for the helpful discussions. The authors thank Anne Dutrey for sharing her PdBI measurements on Haro~6-10.
\end{acknowledgements}

\begin{figure}[h!]
\includegraphics[width=8.3cm]{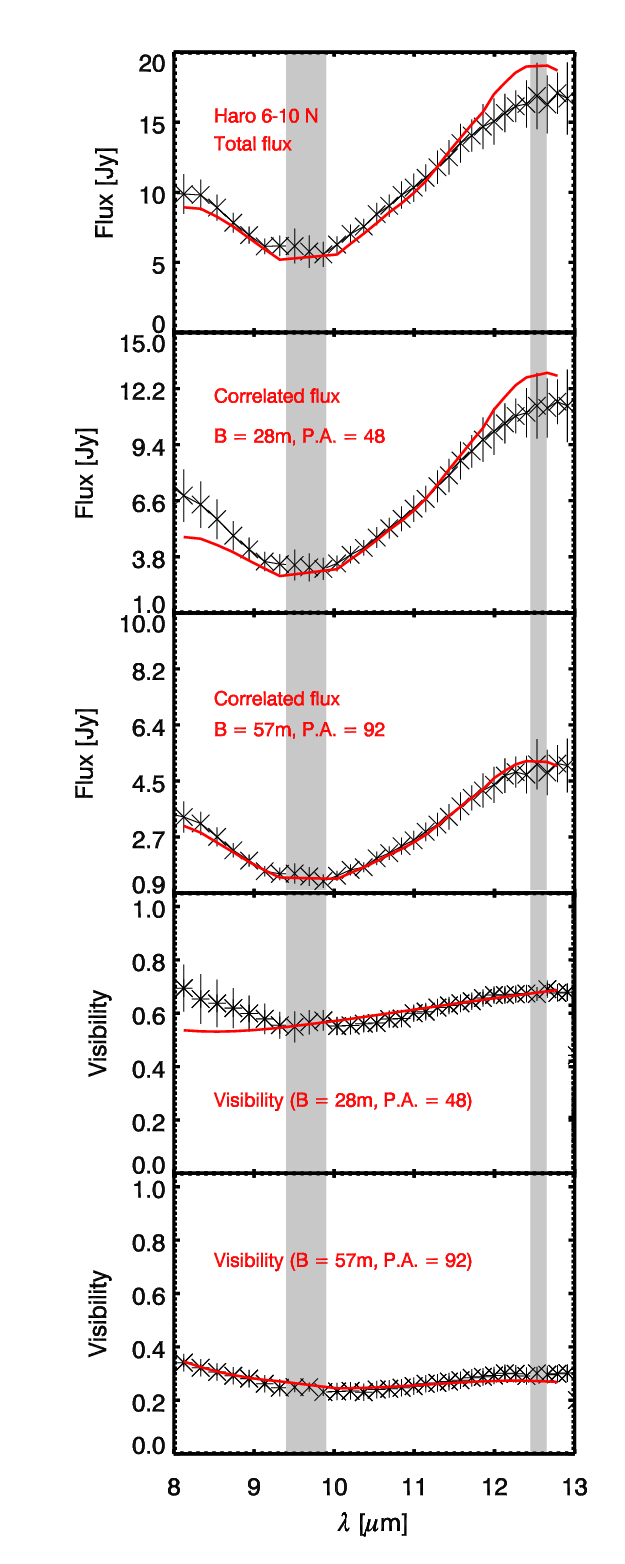}
   \caption{MIDI spectrum, correlated fluxes and visibilities for Haro~6-10~N.
     The red solid line represents the model
     best fit.}
   \label{modelBBG_N}
\end{figure}

\begin{appendix}
\section{Spectral Energy Distribution}

\begin{figure*}
\centering
\includegraphics[width=15cm]{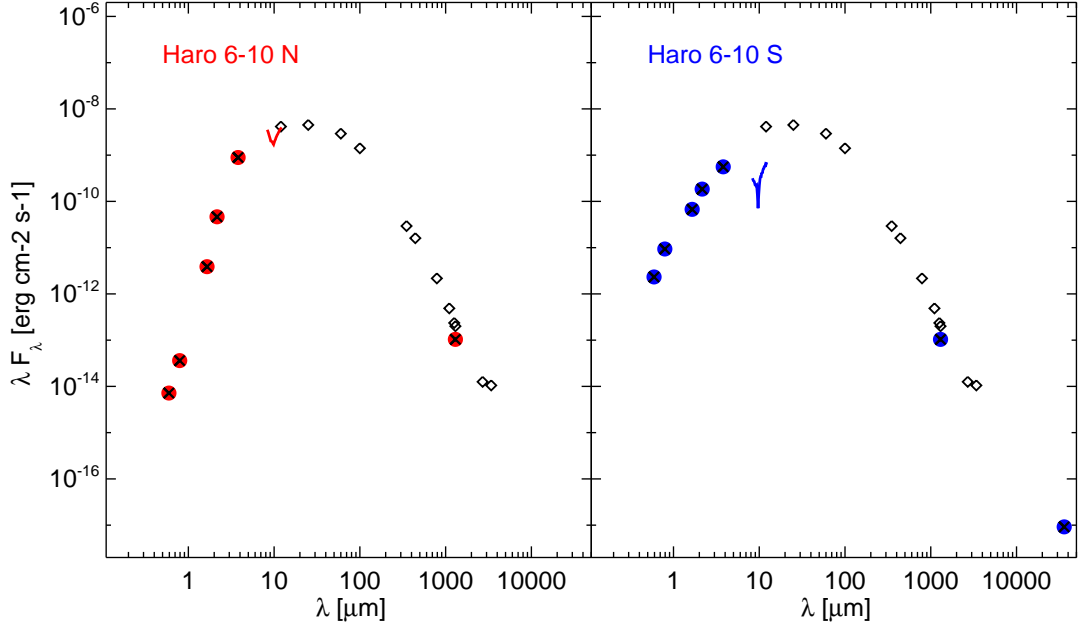}
   \caption{Spectral energy distribution (SED) of Haro~6-10. The empty diamonds
     points represent the photometric measurements of the unresolved
     binary. The fluxes of the single components are presented as red
     (Haro~6-10~N) and blue (Haro~6-10~S) filled dots. The continuous lines
     represent the MIDI spectra of both components.   
           }
      \label{sedgvtau}
\end{figure*}

For completeness, we compile the spectral energy distributions (SEDs) of the two components of the binary system combining our new results in the optical, near- and mid-infrared, with the photometry from the literature. All the fluxes, together with the instruments, the beam sizes of the observations, and the references, are listed in Table~\ref{GVTAU sed}. In Fig.~\ref{sedgvtau} we show the SEDs of the individual components.

In the wavelength range 12-1000~$\mu$m the binary system is not resolved. But the level of the MIDI spectra reveal that most of the far-infrared flux originates in the northern component. 
\begin{table*} 
\caption{Spectral energy distribution with photometry from the literature (references in the last column) of Haro 6-10 N and Haro 6-10~S.}
\label{GVTAU sed}\begin{tabular}{lcccllll}
\hline
\hline
\noalign{\smallskip}
$\lambda$& Flux (N+S) &Flux (S) &Flux (N)&Instrument& Beam size&Date&Reference\\
\noalign{\smallskip}
[$\mu$m]&[mJy]&[mJy]&[mJy]&&[$\arcsec$]&\\
\noalign{\smallskip}
\hline
\noalign{\smallskip}
0.593    & 0.438 $\pm$ 0.001 & 0.437 $\pm$ 0.001    &  0.001 $\pm$ 0.000 & HST        & 0.5 		 & Nov.~1998  & this work \\
0.792    & 2.374 $\pm$ 0.004 & 2.365 $\pm$ 0.004    &  0.009 $\pm$ 0.000 & HST        & 0.5 		 & Nov.~1998  & this work \\
1.650    &  39.2 $\pm$ 3.4   &  37.0 $\pm$ 3.4      &  2.1   $\pm$ 0.2   & NACO       & 0.2 		 & Dec.~2003  & this work \\
2.160    & 166.3 $\pm$ 12.6  & 132.9 $\pm$ 12.2     &  33.4  $\pm$ 3.1   & NACO       & 0.2 		 & Dec.~2003  & this work \\
12       & 16600 $\pm$ 2200  &       --             &   --               & IRAS       &     	         &            & \citet{Chandleretal1998} \\
25       & 37600 $\pm$ 3200  &       --             &   --               & IRAS       &     	         &            & \citet{Chandleretal1998} \\
60       & 58400 $\pm$ 9700  &       --             &   --               & IRAS       &     	         &            & \citet{Chandleretal1998} \\
100      & 47000 $\pm$ 11000 &       --             &   --               & IRAS       &     	         &            & \citet{Chandleretal1998} \\
350      &  3420 $\pm$ 250   &       --             &   --               & JCMT/UKT14 & 19  		 &            & \cite{Dentetal1998}\\
443      &  2370 $\pm$ 500   &       --             &   --               & JCMT/UKT14 & 19  		 & Sep.~1991  & \cite{Chandleretal1998}\\
790      &   571 $\pm$ 44    &       --             &   --               & JCMT/UKT14 & 16  		 & Sep.~1991  & \cite{Chandleretal1998}\\
1104     &   179 $\pm$ 17    &       --             &   --               & JCMT/UKT14 & 19  		 & Sep.~1991  & \cite{Chandleretal1998}\\
1260     &    99 $\pm$ 30    &       --             &   --               & JCMT/UKT14 & 21  		 & Sep.~1991  & \cite{Chandleretal1998}\\
1300     &  90.5 $\pm$ 4.5   & 46.7 $\pm$ 3.2       & 43.8 $\pm$ 3.1     & PdBI       & 0.9$\times$0.6   &            & \cite{Guilloteauetal2011}\\
2700     &  19.6 $\pm$ 1.0   &  9.1 $\pm$ 0.7       & 10.5 $\pm$ 0.7     & PdBI       &                  &            & \cite{Guilloteauetal2011}\\
3400     &  11.9 $\pm$ 1.7   &       --             &   --               & OVRO       & 6                & 1993/1994  & \cite{Hogerheijdeetal1997}\\
36000    &        --         & 1.10                 & 0.10               & VLA        & 0.29$\times$0.26 & March~2002 & \cite{Reipurthetal2004} \\
\noalign{\smallskip}
\hline
\end{tabular}
\end{table*}

\section{Models}
\label{models}
 In the paper the {\em`Two Black-Body Model'} has been used to interpret the MIDI interferometric data.
%
In the following, we list the details of this model.

The 1D Gaussian intensity brightness distribution ($I(\alpha,\beta)$, $\alpha$ and $\beta$ are the coordinates) is:
\begin{equation}\label{I_gauss}
I(\alpha,\beta) = \frac{1}{\sqrt{ \pi /4 ln 2}\Theta}\exp \Big( \frac{-4 \, ln \, 2 \, \rho^2}{\Theta^2}\Big){\rm\ ,}
\end{equation}
where $\Theta$ is the FWHM of the Gaussian, and $\rho=\sqrt{\alpha^2 \, + \, \beta^2}$\ .\\
The Fourier Transform of Eq.~\ref{I_gauss} is:
\begin{equation}\label{eq:Vgauss}
      V(u,v) = \exp \Big(-\frac{( \, \pi \, \Theta \, r  \, )^2}{4 \, ln \, 2}\Big){\rm\ ,}
    \end{equation}
where $r = \sqrt{ \, u^2 \, + \, v^2}$\ .\\

The visibility of a Gaussian brightness distribution inclined by an angle $i$ and rotated by
an angle $\phi$ (which represents the position angle of the major axis) can be obtained by
rotating the $u\,v$-coordinates \citep[e.g., ][]{BergerSegransan2007}:  
  \begin{eqnarray}
      u' & = & u \cdot cos(\phi) + v \cdot sin{\phi}{\rm\ ,} \\
      v' & = & -u \cdot sin(\phi) + v \cdot cos{\phi}{\rm\ ,} \\
\label{eq:uv}
      r  & = &\sqrt{u'^2 + v'^2 \cdot cos(i)^2}{\rm\ .}
    \end{eqnarray}
     Here $cos(i)^2$ represents a compression factor along the minor axis.
      
 The visibility of a 2D Gaussian brightness distribution is given by Eq.~\ref{eq:Vgauss}, but with $r$ defined according to Eq.~\ref{eq:uv}.
    


A circumstellar disc is approximated with two Gaussian components which emit as 
two black-bodies at different temperatures ($T_1, T_2$). Since the infrared spectrum 
is observed in absorption, foreground dust characterised by its optical depth $\tau$ is included.
The black-body emitting at temperature $T$ ($B_\lambda(T)$) and the corresponding total flux received ($F_\lambda(T)$) are:
\begin{eqnarray}
\label{BlT}
B_\nu(T) & = & \frac{2 \cdot h \cdot \nu^3 }{ c^2 } \exp\Big( \frac{h \cdot  \nu}{k \cdot T - 1}\Big){\rm\ ,}\\ 
\label{FlT}
F_\lambda(T) & = & \frac{\pi \cdot  q^2} {d^2} \cdot B_\lambda(T){\rm\ ,}
\end{eqnarray}
where $q$ is the radius of the emitting region (in $AU$) and $d$ is the distance of the source (in $pc$).
For each binary component, the mid-infrared spectrum is then computed as
\begin{equation}
F_{\rm tot}(\lambda) = \Big( F_{\lambda,1}(T_1) \cdot e^{- \tau_1 \cdot
  \frac{\tau(\lambda)}{\tau(\lambda_c)}} + F_{\lambda,2}(T_2)   \cdot e^{- \tau_2 \cdot
  \frac{\tau(\lambda)}{\tau(\lambda_c)}}\Big) \cdot cos(i){\rm\ ,} 
\end{equation}
where $F_{\lambda,1}(T_1)  $ and  $F_{\lambda,2}(T_2) $ are given by Eq.~\ref{FlT}. Since the disc is inclined the received total flux is multiplied by $cos(i)$ \citep[e.g., ][]{BergerSegransan2007}, with $i$ being the inclination of the disc. The function $\tau(\lambda)$ describes the extinction profile between 8 and 13~$\mu$m as measured in the direction of the galactic centre, and $\tau(\lambda_c)$ corresponds to the maximum of $\tau(\lambda)$. $\tau_1$ and $\tau_2$ are the average extinctions computed in Section \ref{anTIMMI2} for Haro~6-10~N and Haro~6-10~S, respectively. 
 The visibility is computed as
\begin{equation}
V_{\rm tot}(\lambda) = f \cdot V_{\rm gauss 1} + (1-f) \cdot V_{\rm gauss 2}{\rm\ ,}\label{Vgauss12}
\end{equation}
where 
$f = \frac{F_{\rm bb 1}}{F_{\rm tot}}$. $V_{\rm gauss 1,2}$ are defined according to Equation~(\ref{eq:Vgauss}), with $\Theta_{1,2}=r_{1,2}/d$, where $r_{1,2}$ is the radius of the emitting region (in $AU$) and $d$ is the distance of the source (in $pc$). Equation~(\ref{Vgauss12}) is valid only, if both $F_{\lambda,1}(T_1)$ and $F_{\lambda,2}(T_2)$ are axisymmetric to the same axis (one going through the star), which is assumed here.

The correlated flux for each component of the binary system is:
\begin{equation}
F_{\rm corr} \, = \, F_{\rm tot}(\lambda) \cdot V_{\rm tot}(\lambda){\rm\ ,} 
\end{equation}
where $F_{\rm tot}(\lambda)$ is given by Eq.~\ref{FlT} and $V_{\rm tot}(\lambda)$ by Eq.~\ref{Vgauss12}.
 
\end{appendix}

\bibliographystyle{aa}
\bibliography{references}

\begin{thebibliography}{70}
\expandafter\ifx\csname natexlab\endcsname\relax\def\natexlab#1{#1}\fi

\bibitem[{{Anglada} {et~al.}(1998){Anglada}, {Villuendas}, {Estalella},
  {Beltr{\'a}n}, {Rodr{\'{\i}}guez}, {Torrelles}, \&
  {Curiel}}]{Angladaetal1998}
{Anglada}, G., {Villuendas}, E., {Estalella}, R., {et~al.} 1998, \aj, 116, 2953

\bibitem[{{Bate} \& {Bonnell}(1997)}]{BateBonnell1997}
{Bate}, M.~R. \& {Bonnell}, I.~A. 1997, \mnras, 285, 33

\bibitem[{{Bate} {et~al.}(2000){Bate}, {Bonnell}, {Clarke}, {Lubow}, {Ogilvie},
  {Pringle}, \& {Tout}}]{Bateetal2000}
{Bate}, M.~R., {Bonnell}, I.~A., {Clarke}, C.~J., {et~al.} 2000, \mnras, 317,
  773

\bibitem[{{Bate} \& {Burkert}(1997)}]{BateBurkert1997}
{Bate}, M.~R. \& {Burkert}, A. 1997, \mnras, 288, 1060

\bibitem[{{Beck} {et~al.}(2010){Beck}, {Bary}, \& {McGregor}}]{Becketal2010}
{Beck}, T.~L., {Bary}, J.~S., \& {McGregor}, P.~J. 2010, \apj, 722, 1360

\bibitem[{{Berger} \& {Segransan}(2007)}]{BergerSegransan2007}
{Berger}, J.~P. \& {Segransan}, D. 2007, \nar, 51, 576

\bibitem[{{Boffin} {et~al.}(1998){Boffin}, {Watkins}, {Bhattal}, {Francis}, \&
  {Whitworth}}]{Boffinetal1998}
{Boffin}, H.~M.~J., {Watkins}, S.~J., {Bhattal}, A.~S., {Francis}, N., \&
  {Whitworth}, A.~P. 1998, \mnras, 300, 1189

\bibitem[{{Bonnell}(1994)}]{Bonnell1994}
{Bonnell}, I.~A. 1994, \mnras, 269, 837

\bibitem[{{Bonnell} \& {Bate}(1994{\natexlab{a}})}]{BonnellBate1994a}
{Bonnell}, I.~A. \& {Bate}, M.~R. 1994{\natexlab{a}}, \mnras, 269, L45

\bibitem[{{Bonnell} \& {Bate}(1994{\natexlab{b}})}]{BonnellBate1994b}
{Bonnell}, I.~A. \& {Bate}, M.~R. 1994{\natexlab{b}}, \mnras, 271, 999

\bibitem[{{Boss} \& {Bodenheimer}(1979)}]{BossBodenheimer1979}
{Boss}, A.~P. \& {Bodenheimer}, P. 1979, \apj, 234, 289

\bibitem[{{Bouwman} {et~al.}(2006){Bouwman}, {Lawson}, {Dominik}, {Feigelson},
  {Henning}, {Tielens}, \& {Waters}}]{Bouwmanetal2006}
{Bouwman}, J., {Lawson}, W.~A., {Dominik}, C., {et~al.} 2006, \apjl, 653, L57

\bibitem[{{Bouwman} {et~al.}(2001){Bouwman}, {Meeus}, {de Koter}, {Hony},
  {Dominik}, \& {Waters}}]{Bouwmanetal2001}
{Bouwman}, J., {Meeus}, G., {de Koter}, A., {et~al.} 2001, \aap, 375, 950

\bibitem[{{Burkert} {et~al.}(1997){Burkert}, {Bate}, \&
  {Bodenheimer}}]{Burkertetal1997}
{Burkert}, A., {Bate}, M.~R., \& {Bodenheimer}, P. 1997, \mnras, 289, 497

\bibitem[{{Chandler} {et~al.}(1998){Chandler}, {Barsony}, \&
  {Moore}}]{Chandleretal1998}
{Chandler}, C.~J., {Barsony}, M., \& {Moore}, T.~J.~T. 1998, \mnras, 299, 789

\bibitem[{{Clarke} \& {Pringle}(1991{\natexlab{a}})}]{ClarkePringle1991b}
{Clarke}, C.~J. \& {Pringle}, J.~E. 1991{\natexlab{a}}, \mnras, 249, 584

\bibitem[{{Clarke} \& {Pringle}(1991{\natexlab{b}})}]{ClarkePringle1991a}
{Clarke}, C.~J. \& {Pringle}, J.~E. 1991{\natexlab{b}}, \mnras, 249, 588

\bibitem[{{Cohen} {et~al.}(1999){Cohen}, {Walker}, {Carter}, {Hammersley},
  {Kidger}, \& {Noguchi}}]{Cohenetal1999}
{Cohen}, M., {Walker}, R.~G., {Carter}, B., {et~al.} 1999, \aj, 117, 1864

\bibitem[{{Davis} {et~al.}(1997){Davis}, {Eisloeffel}, {Ray}, \&
  {Jenness}}]{Davisetal1997}
{Davis}, C.~J., {Eisloeffel}, J., {Ray}, T.~P., \& {Jenness}, T. 1997, \aap,
  324, 1013

\bibitem[{{Davis} {et~al.}(1994){Davis}, {Mundt}, \&
  {Eisloeffel}}]{Davisetal1994}
{Davis}, C.~J., {Mundt}, R., \& {Eisloeffel}, J. 1994, \apjl, 437, L55

\bibitem[{{Dent} {et~al.}(1998){Dent}, {Matthews}, \&
  {Ward-Thompson}}]{Dentetal1998}
{Dent}, W.~R.~F., {Matthews}, H.~E., \& {Ward-Thompson}, D. 1998, \mnras, 301,
  1049

\bibitem[{{Devine} {et~al.}(1999){Devine}, {Reipurth}, {Bally}, \&
  {Balonek}}]{Devineetal1999}
{Devine}, D., {Reipurth}, B., {Bally}, J., \& {Balonek}, T.~J. 1999, \aj, 117,
  2931

\bibitem[{{Doppmann} {et~al.}(2008){Doppmann}, {Najita}, \&
  {Carr}}]{Doppmannetal2008}
{Doppmann}, G.~W., {Najita}, J.~R., \& {Carr}, J.~S. 2008, \apj, 685, 298

\bibitem[{{Duch{\^e}ne} {et~al.}(2007){Duch{\^e}ne}, {Bontemps}, {Bouvier},
  {Andr{\'e}}, {Djupvik}, \& {Ghez}}]{Ducheneetal2007}
{Duch{\^e}ne}, G., {Bontemps}, S., {Bouvier}, J., {et~al.} 2007, \aap, 476, 229

\bibitem[{{Eisl\"offel} {et~al.}(1996){Eisl\"offel}, {Smith}, {Davis}, \&
  {Ray}}]{Eisloffeletal1996}
{Eisl\"offel}, J., {Smith}, M.~D., {Davis}, C.~J., \& {Ray}, T.~P. 1996, \aj,
  112, 2086

\bibitem[{{Fedele} {et~al.}(2008){Fedele}, {van den Ancker}, {Acke}, {van der
  Plas}, {van Boekel}, {Wittkowski}, {Henning}, {Bouwman}, {Meeus}, \&
  {Rafanelli}}]{Fedeleetal2008}
{Fedele}, D., {van den Ancker}, M.~E., {Acke}, B., {et~al.} 2008, \aap, 491,
  809

\bibitem[{{Furlan} {et~al.}(2008){Furlan}, {McClure}, {Calvet}, {Hartmann},
  {D'Alessio}, {Forrest}, {Watson}, {Uchida}, {Sargent}, {Green}, \&
  {Herter}}]{Furlanetal2008}
{Furlan}, E., {McClure}, M., {Calvet}, N., {et~al.} 2008, \apjs, 176, 184

\bibitem[{{Gibb} {et~al.}(2007){Gibb}, {Van Brunt}, {Brittain}, \&
  {Rettig}}]{Gibbetal2007}
{Gibb}, E.~L., {Van Brunt}, K.~A., {Brittain}, S.~D., \& {Rettig}, T.~W. 2007,
  \apj, 660, 1572

\bibitem[{{Guilloteau} {et~al.}(2011){Guilloteau}, {Dutrey}, {Pietu}, \&
  {Boehler}}]{Guilloteauetal2011}
{Guilloteau}, S., {Dutrey}, A., {Pietu}, V., \& {Boehler}, Y. 2011, \aap,
  submitted

\bibitem[{{Hanner} {et~al.}(1998){Hanner}, {Brooke}, \&
  {Tokunaga}}]{Hanneretal1998}
{Hanner}, M.~S., {Brooke}, T.~Y., \& {Tokunaga}, A.~T. 1998, \apj, 502, 871

\bibitem[{{Heller}(1993)}]{Heller1993}
{Heller}, C.~H. 1993, \apj, 408, 337

\bibitem[{{Heller}(1995)}]{Heller1995}
{Heller}, C.~H. 1995, \apj, 455, 252

\bibitem[{{Hogerheijde} {et~al.}(1997){Hogerheijde}, {van Dishoeck}, {Blake},
  \& {van Langevelde}}]{Hogerheijdeetal1997}
{Hogerheijde}, M.~R., {van Dishoeck}, E.~F., {Blake}, G.~A., \& {van
  Langevelde}, H.~J. 1997, \apj, 489, 293

\bibitem[{{Holtzman} {et~al.}(1995){Holtzman}, {Burrows}, {Casertano},
  {Hester}, {Trauger}, {Watson}, \& {Worthey}}]{Holtzmanetal1995}
{Holtzman}, J.~A., {Burrows}, C.~J., {Casertano}, S., {et~al.} 1995, \pasp,
  107, 1065

\bibitem[{{Jaffe} {et~al.}(2004){Jaffe}, {Meisenheimer}, {R{\"o}ttgering},
  {Leinert}, {Richichi}, {Chesneau}, {Fraix-Burnet}, {Glazenborg-Kluttig},
  {Granato}, {Graser}, {Heijligers}, {K{\"o}hler}, {Malbet}, {Miley},
  {Paresce}, {Pel}, {Perrin}, {Przygodda}, {Schoeller}, {Sol}, {Waters},
  {Weigelt}, {Woillez}, \& {de Zeeuw}}]{Jaffeetal2004}
{Jaffe}, W., {Meisenheimer}, K., {R{\"o}ttgering}, H.~J.~A., {et~al.} 2004,
  \nat, 429, 47

\bibitem[{{Jensen} {et~al.}(2004){Jensen}, {Mathieu}, {Donar}, \&
  {Dullighan}}]{Jensenetal2004}
{Jensen}, E.~L.~N., {Mathieu}, R.~D., {Donar}, A.~X., \& {Dullighan}, A. 2004,
  \apj, 600, 789

\bibitem[{{Kemper} {et~al.}(2004){Kemper}, {Vriend}, \&
  {Tielens}}]{Kemperetal2004}
{Kemper}, F., {Vriend}, W.~J., \& {Tielens}, A.~G.~G.~M. 2004, \apj, 609, 826

\bibitem[{{Koresko}(1998)}]{Koresko1998}
{Koresko}, C.~D. 1998, \apjl, 507, L145

\bibitem[{{Koresko} {et~al.}(1999){Koresko}, {Blake}, {Brown}, {Sargent}, \&
  {Koerner}}]{Koreskoetal1999}
{Koresko}, C.~D., {Blake}, G.~A., {Brown}, M.~E., {Sargent}, A.~I., \&
  {Koerner}, D.~W. 1999, \apjl, 525, L49

\bibitem[{{Leinert} {et~al.}(2001){Leinert}, {Beck}, {Ligori}, {Simon},
  {Woitas}, \& {Howell}}]{Leinertetal2001}
{Leinert}, C., {Beck}, T.~L., {Ligori}, S., {et~al.} 2001, \aap, 369, 215

\bibitem[{{Leinert} \& {Haas}(1989)}]{LeinertHaas1989}
{Leinert}, C. \& {Haas}, M. 1989, \apjl, 342, L39

\bibitem[{{Leinert} {et~al.}(1993){Leinert}, {Zinnecker}, {Weitzel},
  {Christou}, {Ridgway}, {Jameson}, {Haas}, \& {Lenzen}}]{Leinertetal1993}
{Leinert}, C., {Zinnecker}, H., {Weitzel}, N., {et~al.} 1993, \aap, 278, 129

\bibitem[{{Mathieu} {et~al.}(2000){Mathieu}, {Ghez}, {Jensen}, \&
  {Simon}}]{Mathieuetal2000}
{Mathieu}, R.~D., {Ghez}, A.~M., {Jensen}, E.~L.~N., \& {Simon}, M. 2000, in
  Protostars and Planets IV, ed. {Mannings, V., Boss, A.P., Russell, S. S.}
  ({Tucson: University of Arizona Press}), 703--+

\bibitem[{{Mathis}(1998)}]{Mathisetal1998}
{Mathis}, J.~S. 1998, \apj, 497, 824

\bibitem[{{Menard} {et~al.}(1993){Menard}, {Monin}, {Angelucci}, \&
  {Rouan}}]{Menardetal1993}
{Menard}, F., {Monin}, J., {Angelucci}, F., \& {Rouan}, D. 1993, \apjl, 414,
  L117

\bibitem[{{Moeckel} \& {Bally}(2006)}]{MoeckelBally2006}
{Moeckel}, N. \& {Bally}, J. 2006, \apj, 653, 437

\bibitem[{{Moeckel} \& {Bally}(2007{\natexlab{a}})}]{MoeckelBally2007b}
{Moeckel}, N. \& {Bally}, J. 2007{\natexlab{a}}, \apjl, 661, L183

\bibitem[{{Moeckel} \& {Bally}(2007{\natexlab{b}})}]{MoeckelBally2007a}
{Moeckel}, N. \& {Bally}, J. 2007{\natexlab{b}}, \apj, 656, 275

\bibitem[{{Momose} {et~al.}(1996){Momose}, {Ohashi}, {Kawabe}, {Hayashi}, \&
  {Nakano}}]{Momoseetal1996}
{Momose}, M., {Ohashi}, N., {Kawabe}, R., {Hayashi}, M., \& {Nakano}, T. 1996,
  \apj, 470, 1001

\bibitem[{{Monin} {et~al.}(1998){Monin}, {Menard}, \&
  {Duchene}}]{Moninetal1998}
{Monin}, J., {Menard}, F., \& {Duchene}, G. 1998, \aap, 339, 113

\bibitem[{{Monin} {et~al.}(2006){Monin}, {M{\'e}nard}, \&
  {Peretto}}]{Moninetal2006}
{Monin}, J., {M{\'e}nard}, F., \& {Peretto}, N. 2006, \aap, 446, 201

\bibitem[{{Movsessian} \& {Magakian}(1999)}]{MovsessianMagakian1999}
{Movsessian}, T.~A. \& {Magakian}, T.~Y. 1999, \aap, 347, 266

\bibitem[{{Olczak} {et~al.}(2010){Olczak}, {Pfalzner}, \&
  {Eckart}}]{Olczaketal2010}
{Olczak}, C., {Pfalzner}, S., \& {Eckart}, A. 2010, \aap, 509, A63+

\bibitem[{{Pringle}(1989)}]{Pringle1989}
{Pringle}, J.~E. 1989, \mnras, 239, 361

\bibitem[{{Przygodda}(2004)}]{przygodda04}
{Przygodda}, F. 2004, PhD thesis, University of Heidelberg

\bibitem[{{Quanz} {et~al.}(2007){Quanz}, {Apai}, \& {Henning}}]{Quanzetal2007}
{Quanz}, S.~P., {Apai}, D., \& {Henning}, T. 2007, \apj, 656, 287

\bibitem[{{Quanz} {et~al.}(2006){Quanz}, {Henning}, {Bouwman}, {Ratzka}, \&
  {Leinert}}]{Quanzetal2006}
{Quanz}, S.~P., {Henning}, T., {Bouwman}, J., {Ratzka}, T., \& {Leinert}, C.
  2006, \apj, 648, 472

\bibitem[{{Ratzka} {et~al.}(2009){Ratzka}, {Schegerer}, {Leinert},
  {{\'A}brah{\'a}m}, {Henning}, {Herbst}, {K{\"o}hler}, {Wolf}, \&
  {Zinnecker}}]{Ratzkaetal2009}
{Ratzka}, T., {Schegerer}, A.~A., {Leinert}, C., {et~al.} 2009, \aap, 502, 623

\bibitem[{{Reipurth} {et~al.}(2004){Reipurth}, {Rodr{\'{\i}}guez}, {Anglada},
  \& {Bally}}]{Reipurthetal2004}
{Reipurth}, B., {Rodr{\'{\i}}guez}, L.~F., {Anglada}, G., \& {Bally}, J. 2004,
  \aj, 127, 1736

\bibitem[{{Sauter} {et~al.}(2009){Sauter}, {Wolf}, {Launhardt}, {Padgett},
  {Stapelfeldt}, {Pinte}, {Duch{\^e}ne}, {M{\'e}nard}, {McCabe}, {Pontoppidan},
  {Dunham}, {Bourke}, \& {Chen}}]{Sauteretal2009}
{Sauter}, J., {Wolf}, S., {Launhardt}, R., {et~al.} 2009, \aap, 505, 1167

\bibitem[{{Schegerer} {et~al.}(2006){Schegerer}, {Wolf}, {Voshchinnikov},
  {Przygodda}, \& {Kessler-Silacci}}]{Schegereretal2006}
{Schegerer}, A., {Wolf}, S., {Voshchinnikov}, N.~V., {Przygodda}, F., \&
  {Kessler-Silacci}, J.~E. 2006, \aap, 456, 535

\bibitem[{{Schegerer} {et~al.}(2009){Schegerer}, {Wolf}, {Hummel}, {Quanz}, \&
  {Richichi}}]{Schegereretal2009}
{Schegerer}, A.~A., {Wolf}, S., {Hummel}, C.~A., {Quanz}, S.~P., \& {Richichi},
  A. 2009, \aap, 502, 367

\bibitem[{{Schegerer} {et~al.}(2008){Schegerer}, {Wolf}, {Ratzka}, \&
  {Leinert}}]{Schegereretal2008}
{Schegerer}, A.~A., {Wolf}, S., {Ratzka}, T., \& {Leinert}, C. 2008, \aap, 478,
  779

\bibitem[{{Stapelfeldt} {et~al.}(1998{\natexlab{a}}){Stapelfeldt}, {Burrows},
  {Krist}, {Watson}, {Ballester}, {Clarke}, {Crisp}, {Evans}, {Gallagher},
  {Griffiths}, {Hester}, {Hoessel}, {Holtzman}, {Mould}, {Scowen}, {Trauger},
  \& {Westphal}}]{Stapelfeldtetal1998}
{Stapelfeldt}, K.~R., {Burrows}, C.~J., {Krist}, J.~E., {et~al.}
  1998{\natexlab{a}}, \apj, 508, 736

\bibitem[{{Stapelfeldt} {et~al.}(1998{\natexlab{b}}){Stapelfeldt}, {Krist},
  {Menard}, {Bouvier}, {Padgett}, \& {Burrows}}]{Stapelfeldtetal1998a}
{Stapelfeldt}, K.~R., {Krist}, J.~E., {Menard}, F., {et~al.}
  1998{\natexlab{b}}, \apjl, 502, L65+

\bibitem[{{Stojimirovi{\'c}} {et~al.}(2007){Stojimirovi{\'c}}, {Narayanan}, \&
  {Snell}}]{Stojimirovicetal2007}
{Stojimirovi{\'c}}, I., {Narayanan}, G., \& {Snell}, R.~L. 2007, \apj, 660, 418

\bibitem[{{van Dokkum}(2001)}]{vanDokkum2001}
{van Dokkum}, P.~G. 2001, \pasp, 113, 1420

\bibitem[{{Wolf} {et~al.}(2003){Wolf}, {Padgett}, \&
  {Stapelfeldt}}]{Wolfetal2003}
{Wolf}, S., {Padgett}, D.~L., \& {Stapelfeldt}, K.~R. 2003, \apj, 588, 373

\bibitem[{{Wolf} {et~al.}(2008){Wolf}, {Schegerer}, {Beuther}, {Padgett}, \&
  {Stapelfeldt}}]{Wolfetal2008}
{Wolf}, S., {Schegerer}, A., {Beuther}, H., {Padgett}, D.~L., \& {Stapelfeldt},
  K.~R. 2008, \apjl, 674, L101

\bibitem[{{Wolf} {et~al.}(2001){Wolf}, {Stecklum}, \& {Henning}}]{Wolfetal2001}
{Wolf}, S., {Stecklum}, B., \& {Henning}, T. 2001, in IAU Symposium, Vol. 200,
  The Formation of Binary Stars, ed. {H.~Zinnecker \& R.~Mathieu}, 295--+

\end{thebibliography}

\end{document}